\documentclass{article}

\usepackage{fontenc, inputenc, calc, indentfirst, fancyhdr, graphicx, epstopdf, lastpage, ifthen, float, amsmath, amssymb, lineno, setspace, enumitem, mathpazo, booktabs, titlesec, etoolbox, tabto, xcolor, colortbl, soul, multirow, microtype, tikz, totcount, changepage, attrib, upgreek, array, tabularx, pbox, ragged2e, tocloft, marginnote, marginfix, enotez, amsthm, 
scrextend, url, geometry, newfloat, caption, seqsplit}
\usepackage{amsmath}
\usepackage{amssymb}
\usepackage{authblk}
\usepackage[sorting=none,url=false,eprint=false,giveninits=true]{biblatex} 
\usepackage{caption}
\usepackage[hidelinks]{hyperref} 
\usepackage{cleveref}
\usepackage{float}

\usepackage{tabularx}
\usepackage{todonotes}
\usepackage[acronym,toc,nogroupskip]{glossaries}
\usepackage{graphicx}
\usepackage{subcaption}
\usepackage{pgfplots}
\usetikzlibrary{patterns,arrows,calc,decorations}
\newacronym{da}{DA}{digital annealing}
\newacronym{fem}{FEM}{finite element method}
\newacronym{minlp}{MINLP}{mixed integer nonlinear programming}
\newacronym{milp}{MILP}{mixed integer linear programming}
\newacronym{ntr-kzfd}{NTR-KZFD}{negative term reductions by Kolmogorov, Zabih, Freedman, and Drineas}
\newacronym{ptr}{PTR}{positive term reductions}
\newacronym{qa}{QA}{quantum annealing}
\newacronym{qubo}{QUBO}{quadratic unconstrained binary optimization}
\newacronym{qpu}{QPU}{quantum processing unit}
\newacronym{sa}{SA}{simulated annealing}
\newacronym{sdk}{SDK}{software developing kit}

\newcommand{\lp}{\left(}
\newcommand{\rp}{\right)}
\newcommand{\lb}{\left[}
\newcommand{\rb}{\right]}

\newcommand{\complianceTensorSymbol}{C}

\newcommand{\complianceTensorComponent}[1]{\complianceTensorSymbol_{#1}}
\newcommand{\stressSymbol}{\sigma}
\newcommand{\stressTensor}{\boldsymbol{\stressSymbol}}
\newcommand{\stressTensorComponent}[1]{\stressSymbol_{#1}}
\newcommand{\stress}{\stressSymbol}
\newcommand{\strainSymbol}{\varepsilon}

\newcommand{\strainTensorComponent}[1]{\strainSymbol_{#1}}
\newcommand{\displacementSymbol}{u}
\newcommand{\displacement}{\displacementSymbol}
\newcommand{\displacementPrescribed}{\hat{\displacement}}

\newcommand{\tractionSymbol}{t}
\newcommand{\traction}{\tractionSymbol}

\newcommand{\tractionPrescribed}{\hat{\traction}}

\newcommand{\bodyForceDensity}{f}

\newcommand{\domain}{\Omega}
\newcommand{\boundary}{\Gamma}
\newcommand{\boundaryDisp}{\boundary^{\displacement}}
\newcommand{\boundaryTraction}{\boundary^{\stressSymbol}}
\newcommand{\normal}{n}

\newcommand{\intDomain}[1]{\int_{\domain} #1\,\text{d}\domain}
\newcommand{\intBoundaryDisp}[1]{\int_{\boundaryDisp} #1\,\text{d}\boundary}
\newcommand{\superscriptComplementaryEnergy}{*}
\newcommand{\complStrainEnergy}{U^{\superscriptComplementaryEnergy}}
\newcommand{\complStrainEnergyDensity}{\complStrainEnergy_0}
\newcommand{\complExternal}{W^{\superscriptComplementaryEnergy}}
\newcommand{\complExternalVolume}{W^{\superscriptComplementaryEnergy V}}
\newcommand{\complExternalVolumeDensity}{\complExternalVolume_0}
\newcommand{\complExternalSurface}{W^{\superscriptComplementaryEnergy S}}
\newcommand{\complExternalSurfaceDensity}{\complExternalSurface_0}
\newcommand{\totalComplEnergy}{\Pi^{\superscriptComplementaryEnergy}}

\newcommand{\setStaticallyAdmissible}{\boldsymbol{\mathcal{S}}}

\newcommand{\designVarSymbol}{\alpha}
\newcommand{\designVarVector}{\boldsymbol{\designVarSymbol}}
\newcommand{\designVarComponent}[1]{\designVarSymbol_{#1}}
\newcommand{\setAdmissibleDesign}{\boldsymbol{\mathcal{A}}}
\newcommand{\intElement}[2]{\int_{x_{#1}}^{x_{#1+1}} #2\,\text{d}x}

\newcommand{\rodLength}{L}

\newcommand{\youngModulus}[1][]{E_{#1}}
\newcommand{\crossSectionalArea}[1][]{A_{#1}}
\newcommand{\x}{x}
\newcommand{\elemIndex}{e}
\newcommand{\nElem}{n_e}
\newcommand{\nodeIndex}{i}

\newcommand{\basisFctSymbol}{\phi}
\newcommand{\basisFct}{\basisFctSymbol}
\newcommand{\basisCoeff}{a}
\newcommand{\forceSymbol}{F}
\newcommand{\force}{\forceSymbol}
\newcommand{\forceAnalytic}{\forceSymbol^{*}}

\newcommand{\nQubitsPerNode}{n_q}
\newcommand{\nQubitsInput}{N_q}
\newcommand{\nQubitsLogical}{\hat{N}_q}
\newcommand{\qubitSymbol}{q}
\newcommand{\qubitsVector}{\boldsymbol{\qubitSymbol}}
\newcommand{\qubitCoeff}{\qubitSymbol^{\basisCoeff}}
\newcommand{\qubitsVectorCoeffs}{\qubitsVector^{\basisCoeff}}
\newcommand{\qubitDesign}{\qubitSymbol^{\crossSectionalArea}}
\newcommand{\qubitsVectorDesigns}{\qubitsVector^{\crossSectionalArea}}

\newcommand{\objective}{J}
\newcommand{\penaltyWeight}{\lambda}
\newcommand{\penaltyTerm}{\pi}

\newcommand{\crossSectionalAreaChoice}[1]{\mathcal{A}_{#1}}
\newcommand{\objectiveQuadratic}{\Hat{\objective}}
\newcommand{\qubitsVectorAuxiliary}{\boldsymbol{\Hat{\qubitSymbol}}}

\newcommand{\tAnnealing}{t_{A}}

\newcommand{\nreads}{n_{\mathrm{reads}}}

\title{A Formulation of Structural Design Optimization Problems for Quantum Annealing}

\author[1,*]{Fabian Key}
\author[1]{Lukas Freinberger}
\affil[1]{Institute of Lightweight Design and Structural Biomechanics (ILSB), TU Wien, Vienna, Austria}

\begin{document}
\maketitle


\begin{abstract}
    We present a novel formulation of structural design optimization problems specifically tailored to be solved by \gls{qa}. Structural design optimization aims to find the best, i.e., material-efficient yet high-performance, configuration of a structure. To this end, computational optimization strategies can be employed, where a recently evolving strategy based on quantum mechanical effects is \gls{qa}. This approach requires the optimization problem to be present, e.g., as a \gls{qubo} model. Thus, we develop a novel formulation of the optimization problem. The latter typically involves an analysis model for the component. Here, we use energy minimization principles that govern the behavior of structures under applied loads. This allows us to state the optimization problem as one overall minimization problem. Next, we map this to a \gls{qubo} problem that can be immediately solved by \gls{qa}. 
    We validate the proposed approach using a size optimization problem of a compound rod under self-weight loading. 
    To this end, we develop strategies to account for the limitations of currently available hardware and find that
    the presented formulation is suitable for solving structural design optimization problems through \gls{qa} and, for small-scale problems, already works on today's hardware.
\end{abstract}



\glsresetall

\section{Introduction}
\label{sec:introduction}
Could the emergence of \gls{qa}, a computational optimization approach leveraging quantum mechanical effects, make an investigation of new strategies for structural design optimization worthwhile? In our pursuit of unraveling an answer, we address the particular requirements and advantages of \gls{qa} in a novel formulation of structural design optimization problems.
In such a problem, one seeks the optimal configuration of a structure concerning performance criteria, resource utilization, and engineering constraints. By optimizing structural designs, engineers can create safer, more efficient, and environmentally friendly components and structures. However, modern engineering challenges often involve complex design spaces with numerous variables, constraints, and objectives. These spaces may be difficult to explore using traditional design methods. For example, these methods may struggle with problems involving a large number of local optima or an exponential growth of solution possibilities. This is where it becomes interesting to examine the capabilities of \gls{qa}. If the above problems are formulated in a specific way, e.g., as \gls{qubo} problems, \gls{qa} has the potential to work more efficiently on these types of problems and may be able to find better solutions. Thus, we explore the potential of \gls{qa} for structural design optimization in this work.
\bigskip\par
Structural design optimization encompasses several approaches. One is \textit{topology optimization} in which one is concerned with finding the optimal arrangement of material in a given space. In \textit{shape optimization}, the goal is to find the optimal geometric shape of a structural component. When the shape is fixed, but optimal dimensions of one or more components are sought, a \textit{size optimization} problem is considered. Finally, \textit{material optimization} explores the selection of materials with specific properties to optimize the characteristics of a structure.
\par
Nowadays, these problems are typically solved using computational methods that involve two key components: an optimization algorithm to explore the design space and an analysis model to simulate the behavior of the design. For the latter, structural design optimization often relies on the \gls{fem}.
\par
In contrast, the purpose of this work is to present and evaluate a novel approach that does not treat the optimization algorithm and the analysis model separately, but uses \gls{qa} to find the optimal design and compute the structural response of a component in one go. For the first time, we show that a structural design optimization problem, once properly formulated, can be solved solely using \gls{qa} on currently available quantum hardware.
\bigskip\par
\gls{qa} has already been used to address a range of challenges across multiple scientific disciplines. A comprehensive description covering applications from mobility, scheduling and logistics, quantum simulation, machine learning, and finance can be found in~\cite{Yarkoni2022}. In the field of structural mechanics, however, the existing literature is still quite limited~\cite{TostiBalducci2022}. In one contribution~\cite{Srivastava2019}, the deformation of a rod under axial loading is computed by an iterative algorithm that minimizes a discretized version of the system's potential energy at each step. 
\par
Regarding design optimization, some work related to \gls{qa} has been presented recently. A shape optimization problem for the reduction of noise has been considered in~\cite{Neukart2019}. The authors discuss results for optimizing the shape of a three-dimensional body in an acoustic scattering problem. The respective analysis model was based on the classical \gls{fem} but optimization tasks were iteratively solved using \gls{qa}. 
\par
In the field of computational electromagnetism, the optimization of a planar magnet array~\cite{Maruo2020} and topology optimization of a three-dimensional permanent magnet~\cite{Maruo2022} have been conducted. The latter was also based on an iterative procedure using \gls{fem} for the magneto-static field computations. Both works, however, did not use actual \gls{qa} hardware but a \gls{da} engine instead. 
\par
The work in~\cite{Matsumori2022} presents an approach for design optimization of a printed circuit board. The optimization goal was to use a minimum number of mounting holes while avoiding resonance. Since the related frequency analysis based on the \gls{fem} could not be expressed directly as \gls{qubo} problem, a machine learning technique was used to construct an appropriate model. Results have been presented for a random search approach, \gls{sa}, and hybrid \gls{qa}, an approach that combines \gls{qa} and classical algorithms. 
In a similar way, \gls{qa} has been integrated into a black-box optimization approach to find optimal designs of a noise filter~\cite{Okada2023}. 
In this approach, \gls{qa} is used to solve regression problems appearing in the sequential learning method under consideration, which relies on the \gls{fem} for data acquisition. The results presented include a comparison of the design obtained by incorporating \gls{qa}, \gls{sa}, or random search.
\par
In another study~\cite{Ye2023}, the authors have performed a topology optimization for a minimum compliance problem in which they consider a rectangular domain with a unit point force acting on it. To that end, they transform the original \gls{minlp} problem into a sequence of \gls{milp} problems by separating field and design variables that are updated in an iterative manner. In each step, the field variables are the solution of the linear system resulting from the analysis model in use, namely the \gls{fem}. The \gls{milp} problem for the design variables, on the other hand, is converted to \gls{qubo} form. Since the resulting \gls{qubo} problem involves all-to-all interactions and may become too large, another splitting is performed: one part of the problem is solved classically and for the other part \gls{qa} and hybrid \gls{qa} are used to tackle a reduced \gls{qubo} problem.  
\par
Finally, the optimization of two-dimensional truss structures has been addressed in~\cite{Wils2023} as size optimization problems. The approach aims to find the optimal distribution of cross-sectional areas per truss element to minimize the difference between the truss element stress and the maximum limit stress. Here, the stress evaluation is based on a symbolic \gls{fem}, a bottleneck of the approach as the authors state. The presented method further requires an iterative procedure to transform the resulting fractional objective function into \gls{qubo} form. In this form, all-to-all interactions occur, which has been tackled by a complexity reduction through removal and truncation of specific terms. That allowed to present optimization results for examples that range from two- to four-truss systems.  
\bigskip\par
In distinction to the former methods, which rely on the \gls{fem} as an analysis model, the novel formulation proposed in this work integrates the respective governing equations directly into the optimization problem. The crucial point is that we compute the response of the structural system under applied loads through the principle of minimum complementary energy. The minimization character of this approach allows us to combine it with the objective of the optimization, which is typically also formulated as minimization problem. To this end, we will consider a minimum compliance problem, a common use case in structural design optimization. The resulting overall minimization problem can be formulated as a \gls{qubo} problem just using existing strategies such as degree reduction. The solution of the \gls{qubo} problem can then be obtained by \gls{qa} and represents both the optimal design and the response of the structure. It is worthwhile to emphasize that, in contrast to the referenced works,  the presented method does not include any iterative procedure but only requires the solution of a single \gls{qubo} problem, although the probabilistic nature of the approach demands a certain number of evaluations. Furthermore, the formulation leads to an advantageous structure of the problem, i.e., it does not contain any all-to-all connections between its variables, but only limited local interactions. 
\par
To illustrate the aptitude of the approach, we apply it to a size optimization problem for a rod under self-weight loading. The rod will consist of multiple components, with the respective cross-sectional areas being the design choices. Here, the complexity of the example is adapted to the currently available \gls{qa} hardware. At the moment, the hardware clearly limits the scale of feasible problems. Nevertheless, the presented concept itself is fundamental enough to serve as a new paradigm in structural design optimization using \gls{qa} and, thus, worthwhile to be studied in parallel to the ongoing development of hardware. 
\bigskip\par
In the remainder of this work, we will first present the formulation of the minimum compliance problem in combination with the principle of minimum complementary energy in \Cref{subsec:principleMinimumComplementaryEnergy,subsec:structuralDesignOptimizationProblem}. In a subsequent step, we transform the resulting overall minimization problem into a \gls{qubo} problem suitable for \gls{qa} (see \Cref{subsec:problemFormulation1DRod}). To validate the approach, \Cref{sec:results} presents results from numerical experiments for a rod under self-weight loading, which is composed of multiple elements. Finally, we provide a discussion of the results in \Cref{sec:discussion}.
\section{Materials and Methods}
\label{sec:materialsAndMethods}
In the following, we will derive the \gls{qubo} formulation of the structural design optimization problem. First, we discuss the principle of minimum complementary energy and state the pure structural analysis problem, i.e., for a fixed design. In the next step, we extend this problem by additional design variables and introduce the minimum compliance problem as a structural design optimization problem. In this context, we consider a size optimization problem for a rod under self-weight loading. For both the analysis and the design optimization problems, we introduce a finite-dimensional ansatz that leads to the corresponding \gls{qubo} formulations.
\subsection{The Principle of Minimum Complementary Energy}
\label{subsec:principleMinimumComplementaryEnergy}
The \textit{principle of minimum complementary energy} has a long tradition in solid mechanics. It can be traced back to work presented in \cite{Engesser1889} and \cite{Westergaard1942}. Detailed descriptions can also be found in current textbooks \cite{Reddy2017,Mang2018}, which  serve as the basis for the following explanation. The principle can be used to predict how a structure will behave under external loads. We refer to this as the \textit{structural analysis problem}. 
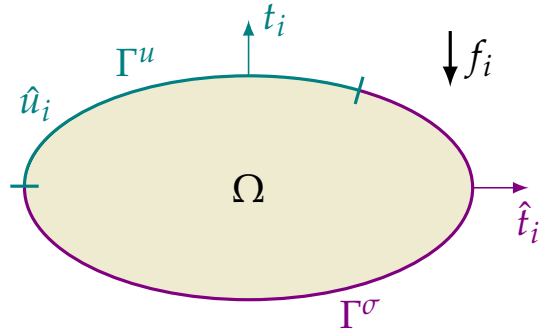
\begin{figure}
    \centering
    \resizebox{0.5\textwidth}{!}{
        \begin{tikzpicture}
            \draw [fill =olive!15] (0,0) ellipse (2 and 1);
            \node at (0,0) {$\Omega$};
            \draw[thick, teal, |-|] (-2,0) node [shift=({0.1,0.75})]{$\displacementPrescribed_i$} arc (-180:-300:2 and 1) node [midway, above] {$\boundaryDisp$};
            \draw [teal,-latex] (0,1) --++ (0, 0.5) node [right, teal] {$\traction_i$};
            \draw[thick, violet] (-2,0) arc (180:420:2 and 1) node [midway, below] {$\boundaryTraction$};
            \draw[violet, -latex](2,0) --++ (0.5,0) node [below] {$\tractionPrescribed_i$};
            \draw[thick, -latex] (1.8,1.4) --++ (0,-0.5) node [midway, right] {$\bodyForceDensity_i$};
        \end{tikzpicture}
    }
    \caption{A generic elastic body $\Omega$ under external loading with prescribed surface traction $\tractionPrescribed_i$ and displacement $\displacementPrescribed_i$ on the boundary portions $\boundaryTraction$ and $\boundaryDisp$, respectively, and body force density $\bodyForceDensity_i$.}
    \label{fig:setupElasticBody}
\end{figure}

To illustrate the usage of the principle, we consider a generic elastic body given in the domain $\domain$ (cf. \Cref{fig:setupElasticBody}). The principle is formulated with respect to the structure's stress field $\stressTensorComponent{ij}$, where $\traction_i=\stressTensorComponent{ij}\normal_j$ is the traction for a unit normal vector $\normal_j$. The displacement is given as $\displacement_i$ and a body force density $\bodyForceDensity_i$ may act on the entire domain. The boundary of the body is denoted by $\boundary$, where surface traction $\tractionPrescribed_i$ and displacement $\displacementPrescribed_i$ are prescribed on the portions $\boundaryTraction$ and $\boundaryDisp$, respectively. 
\bigskip\par
The principle's field of application results from the underlying assumptions in its derivation. These assumptions will be outlined next. First, the principle applies to the response of an \textit{elastic} body due to a specific load that, in our case, will be \textit{static}. Furthermore, we assume that only \textit{small deformations} and \textit{infinitesimal strains} occur. A central condition is the \textit{existence of potentials}. Thus, we assume that a \textit{complementary energy potential} or \textit{complementary strain energy density} $\complStrainEnergyDensity$ exists such that the strain tensor $\strainTensorComponent{ij}$ is given by
\begin{equation}
    \strainTensorComponent{ij} = \frac{\partial\complStrainEnergyDensity}{\partial\stressTensorComponent{ij}}.
\end{equation}
As an additional consequence, we only consider \textit{conservative} external loads, i.e., loads that can be derived from corresponding potentials for volumetric and surface forces. They are denoted as $\complExternalVolumeDensity$ and $\complExternalSurfaceDensity$, respectively.
\bigskip\par
Based on this, we can define the \textit{internal} and the \textit{external complementary energy}, referred to as $\complStrainEnergy$ and $\complExternal$, respectively. The latter contains volumetric and surface-related contributions $\complExternalVolume$ and $\complExternalSurface$, i.e., $\complExternal=\complExternalVolume+\complExternalSurface$. These quantities are given as:
\begin{equation}
    \complStrainEnergy = \intDomain{\complStrainEnergyDensity}, \quad
    \complExternalVolume = \intDomain{\complExternalVolumeDensity}, \quad
    \complExternalSurface = \intBoundaryDisp{\complExternalSurfaceDensity}.    
\end{equation}
The \textit{total complementary energy} $\totalComplEnergy$ follows as
\begin{equation}
    \totalComplEnergy
    =
    \complStrainEnergy
    +
    \complExternal.
\end{equation}
\par
Before we can state the final principle, we have to consider only stress fields $\stressTensorComponent{ij}$ that are \textit{statically admissible}, i.e., they are in  \textit{equilibrium} with a given body force density $\bodyForceDensity_{i}$ in $\domain$ and satisfy the \textit{traction boundary conditions} on $\boundaryTraction$:
\begin{align}
    \stressTensor_{ij,j}
    + \bodyForceDensity_{i}
    &=
    0
    \quad\text{in } \domain,
    \\
    \stressTensorComponent{ij}\normal_{j}
    &=
    \tractionPrescribed_i
    \quad\text{on } \boundaryTraction.
\end{align}
\par
Then it can be shown that in this case the total complementary energy takes a stationary value:
\begin{equation}
    \delta\totalComplEnergy
    =
    \delta\left(
        \complStrainEnergy
        + \complExternal
    \right)
    = 0,
\end{equation}
and we can state the principle of minimum complementary energy, which is formulated in 
\cite{Hoff1956} (Theorem 15b) as follows:
\begin{quote}
    The total complementary potential is a minimum with respect to variations in stress when the system is in its true state of equilibrium.
\end{quote}
So, we consider the total complementary energy $\totalComplEnergy$ as as a functional of the stresses $\stressTensor$, that is
\begin{equation}
    \totalComplEnergy\left[\stressTensor\right]
    =
    \complStrainEnergy\left[\stressTensor\right]
    +
    \complExternal\left[\stressTensor\right],
\end{equation}
and formulate the structural analysis problem as a minimization problem:
\begin{equation}
    \min_{\stressTensor\in\setStaticallyAdmissible}\left\{
    \totalComplEnergy\left[\stressTensor\right]
    \right\},
    \label{eq:structuralAnalysisProblem}
\end{equation}
where $\setStaticallyAdmissible$ denotes the set of all statically admissible stress fields. Consequently, the optimal solution $\stressTensor_{\text{opt}}$ of \Cref{eq:structuralAnalysisProblem} will be the one that corresponds to the actual equilibrium configuration of the elastic body.
\par
Although the principle itself is not restricted to the linear elastic case, for the remainder of this article we will use the generalized Hooke’s law, such that the internal complementary energy reads
\begin{equation}
    \complStrainEnergy
    =
    \frac{1}{2}
    \intDomain{%
        \complianceTensorComponent{ijkl}
        \stressTensorComponent{ij}
        \stressTensorComponent{kl}
    },
    \label{eq:complStrainEnergy}
\end{equation}
where $\complianceTensorComponent{ijkl}$ is the \textit{compliance tensor}.
\par
For completeness, we also provide the expressions used for the external complementary energy:
\begin{equation}
    \complExternalVolume
    =
    - \intDomain{%
        \displacement_i
        \bodyForceDensity_i
    },
    \quad
    \complExternalSurface
    =
    - \intBoundaryDisp{%
        \displacementPrescribed_i
        \traction_i
    }.
\end{equation}
\subsection{The Structural Design Optimization Problem}
\label{subsec:structuralDesignOptimizationProblem}
In structural design optimization, one is interested in finding the best design that is effective --- in the sense that it meets specific performance criteria --- while minimizing cost and resource utilization within given constraints. As a well-known example for this type of problem, we consider the \textit{minimum compliance} problem in the following. By minimizing a component's compliance, its stiffness is inherently maximized, which means it will resist deformation and deflection more effectively under applied loads. 
\bigskip\par
To assess the compliance, it is necessary to solve a structural analysis problem, such as the one described in \Cref{subsec:principleMinimumComplementaryEnergy}. While the solution to this problem provides information about the performance of the design, the resource usage and constraints, on the other hand, are related to the design itself. To that end, \textit{design variables} $\designVarComponent{i}$ and a \textit{set of admissible design variables} $\setAdmissibleDesign$ are introduced. For now, we consider the design variables $\designVarComponent{i}$ to be generic, so that different types of design optimization approaches, as described in \Cref{sec:introduction}, are covered. For example, these variables may describe a distribution of material in a topology optimization problem. In another scenario, $\designVarVector$ may instead refer to a choice of component dimensions or material when dealing with a size or material optimization problem, respectively.
\bigskip\par
The final structural design optimization problem is obtained by extending the structural analysis problem from \Cref{eq:structuralAnalysisProblem} to include the design variables $\designVarVector$. To indicate that the total complementary energy is no longer just a functional of the stresses $\stressTensor$, but additionally parametric in the design variables, we write $\totalComplEnergy\left[\stressTensor;\designVarVector\right]$.
So, the minimum compliance problem can be stated as follows (cf. ~\cite{Bendsoe2004} Eq. (1.7)):
\begin{equation}
    \min_{\designVarVector\in\setAdmissibleDesign}\min_{\stressTensor\in\setStaticallyAdmissible}
    \left\{
        \totalComplEnergy\left[\stressTensor;\designVarVector\right]
    \right\}.
    \label{eq:designOptimizationProblem}
\end{equation}
The solution of this problem $\{\stressTensor_{\text{opt}},\designVarVector_{\text{opt}}\}$ then yields both the true stress field and the best design, i.e., the design with minimal compliance. 
\subsection{Problem Formulations for a Rod under Self-Weight Loading}
\label{subsec:problemFormulation1DRod}
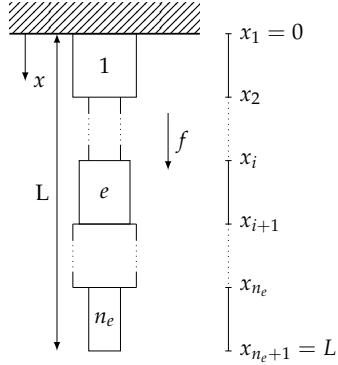
\begin{figure}
    \centering\resizebox{0.3\textwidth}{!}{%
    \begin{tikzpicture}
        \fill[pattern=north east lines] (0,0) rectangle (3,0.5);
        \draw[thick] (0,0) -- (3,0);
        \draw[-latex] (0.25,0) -- (0.25, -0.75) node [right] {$\x$};
        \draw[latex-latex] (0.75,0) -- node[left]{\rodLength}(0.75,-5);
        
        \newcommand{\xa}{3.4}
        \newcommand{\xb}{3.5}
        \newcommand{\xm}{3.45}
        \draw (1,0) rectangle (2,-1);
        \node at (1.5,-0.5) {$1$};
        \draw (\xa,-0) -- (\xb,-0) node [right] {$\x_1=0$};
        \draw (\xa,-1) -- (\xb,-1) node [right] {$\x_2$};
        \draw (1.25,-1) -- (1.25,-1.25);
        \draw (1.75,-1) -- (1.75,-1.25);
        \draw [dotted] (1.25,-1.25) -- (1.25, -1.5);
        \draw [dotted] (1.75,-1.25) -- (1.75, -1.5);

        \draw[dotted] (1.25,-1.5) -- (1.25,-1.75);
        \draw[dotted] (1.75,-1.5) -- (1.75,-1.75);
        \draw (1.25,-1.75) -- (1.25,-2);
        \draw (1.75,-1.75) -- (1.75,-2);
        \draw (\xa,-2) -- (\xb,-2) node [right] {$\x_{\nodeIndex}$};
        \draw (1.1,-2) rectangle (1.9,-3);
        \node at (1.5,-2.5) {$\elemIndex$};
        \draw (\xa,-3) -- (\xb,-3) node [right] {$\x_{\nodeIndex+1}$};
        \draw (1,-3) -- (2,-3);
        \draw (1,-3) -- (1,-3.25);
        \draw (2,-3) -- (2,-3.25);
        \draw[dotted] (1,-3.25) -- (1,-3.5);
        \draw[dotted] (2,-3.25) -- (2,-3.5);
        \draw[dotted] (1,-3.5) -- (1,-3.75);
        \draw[dotted] (2,-3.5) -- (2,-3.75);
        \draw (1,-3.75) -- (1,-4);
        \draw (2,-3.75) -- (2,-4);
        \draw (1,-4) -- (2,-4);
        \draw (\xa,-4) -- (\xb,-4) node [right] {$\x_{\nElem}$};
        \draw (1.25,-4) rectangle (1.75,-5);
        \node at (1.5,-4.5) {$\nElem$};
    
        \draw (\xa,-5) -- (\xb,-5) node [right] {$\x_{\nElem+1}=\rodLength$};
        \draw (\xm,0) -- (\xm,-1); 
        \draw[dotted] (\xm,-1) -- (\xm,-2);
        \draw (\xm,-2) -- (\xm,-3);
        \draw[dotted] (\xm,-3) -- (\xm,-4);
        \draw (\xm,-4) -- (\xm,-5);
        \draw[-latex] (2.5, -1.25) -- node [right] {$\bodyForceDensity$}(2.5, -2.15) ;
  
    \end{tikzpicture}
    }

    \caption{Generic setup for a rod under self-weight loading that is composed of multiple elements $\elemIndex$.}
    \label{fig:composedRod}
\end{figure}

Next, we will consider a specific problem to lay the foundation for formulating the \gls{qubo} problems in \Cref{subsec:qubo1DRod}. To ensure that the resulting problem complexity is appropriate for existing \gls{qa} hardware, we use a rather simple example here. In particular, we consider the one-dimensional problem for a rod of \textit{length} $\rodLength$ under self-weight loading via the body force density $\bodyForceDensity$. The rod will be composed of several \textit{elements} $\elemIndex=1,\dots,\nElem$ spanning from $\x_{\nodeIndex}$ to $\x_{\nodeIndex+1}$, where $\nodeIndex=1,\dots,\nElem+1$ represents the interfaces or, from a purely one-dimensional point of view, nodes between elements. \Cref{fig:composedRod} shows an example setup for illustrative purposes.
Each element can have an individual \textit{cross-sectional area} $\crossSectionalArea[\elemIndex]$ and \textit{Young's modulus} $\youngModulus[\elemIndex]$. Stress and traction related to each element  are denoted by $\stress_{\elemIndex}(\x)$ and $\traction_{\elemIndex}(\x)$, respectively. In addition, we refer to $\displacement_{\elemIndex}$ as the corresponding displacement. 
\bigskip\par
For a better understanding of the notation below, note that from an element point of view, the corresponding nodes are unique, i.e., $\elemIndex \Rightarrow \left\{\nodeIndex, \nodeIndex+1\right\}$ and we can use $\elemIndex$ and $\nodeIndex$ interchangeably. However, the reverse is not true, since one node is part of two elements, so $\nodeIndex \nRightarrow \elemIndex$.
\bigskip\par
We begin with a brief description of the boundary and coupling conditions considered in this test case.
At the upper boundary ($\x=0$), the rod is fixed such that $\displacementPrescribed\rvert_{\x=0} = 0$. For the opposite, free-hanging boundary we prescribe zero traction $\tractionPrescribed\rvert_{\x=\rodLength}=0$.
At each interface between two elements $\elemIndex-1$ and $\elemIndex$, we require the equilibrium of forces: 
\begin{equation}
    \traction_{\elemIndex-1}
    \lp \x_{\elemIndex}\rp \crossSectionalArea[\elemIndex-1] 
    =
    -
    \traction_{\elemIndex}
    \lp \x_{\elemIndex}\rp \crossSectionalArea[\elemIndex]
    .
    \label{eq:equilibriumForces}
\end{equation}
\par
In the following, we consider each element $\elemIndex$ separately and couple adjacent elements through the force equilibrium condition stated above. 
Thus, let us explain the resulting individual boundary conditions for each element.
For the first element ($\elemIndex=1$), we set the displacement at its boundary located at $\x_1=0$ to zero, that is
\begin{equation}
    \displacementPrescribed_1\rvert_{\x=0} = 0.
\end{equation}
Each element is coupled to its predecessor through \Cref{eq:equilibriumForces}. Since the first element has no predecessor, no additional condition needs to be considered here.
On the contrary, all other elements are coupled through \Cref{eq:equilibriumForces}, which results in corresponding traction boundary conditions
\begin{equation}
 \tractionPrescribed_{\elemIndex}\rvert_{\x=\x_{\elemIndex}} 
 = 
 - \traction_{\elemIndex-1}\rvert_{\x=\x_{\elemIndex}}\frac{\crossSectionalArea[\elemIndex-1]}{\crossSectionalArea[\elemIndex] },
 \quad
 \elemIndex=2,\dots,\nElem
 .
 \label{eq:couplingConditions}
\end{equation}
Finally, for the last element $\nElem$ we additionally demand that 
\begin{equation}
 \tractionPrescribed_{\nElem}\rvert_{\x=\rodLength}=0, 
 \label{eq:noTractionBC}
\end{equation}
due to the zero-traction boundary condition at the free-hanging end.
\bigskip\par
Next, we will derive the expression for the total complementary energy that we will minimize with respect to $\stress_{\elemIndex}$ later. The internal complementary strain energy (cf. \Cref{eq:complStrainEnergy}) is given in terms of the element-specific stresses $\stress_{\elemIndex}$ by
\begin{equation}
    \complStrainEnergy_\elemIndex
    \left[\stress_{\elemIndex}\right]
    =
    \frac{1}{2}
    \intElement
    {\elemIndex}
    {
        \frac{1}{\youngModulus[\elemIndex]}
        \crossSectionalArea[\elemIndex]
        \stress_{\elemIndex}^2\left(\x\right)
    }.
    \label{eq:internalComplStrainEnergyElement}
\end{equation}
For the external complementary energy in each element $\complExternal_\elemIndex$, we have on the one hand the element-wise volumetric contributions of $\complExternalVolume$, i.e., $\complExternalVolume_{\elemIndex}$. On the other hand, we also have the surface-related external complementary energy $\complExternalSurface$. It is defined over $\boundaryDisp$ and, thus, only relevant for element 1, where the displacement is prescribed at $\x=0$. Nevertheless, we have $\complExternalSurface_1=0$ in this case, since $\displacementPrescribed\rvert_{x=0}=0$. For the total complementary energy, it follows that
\begin{equation}
    \totalComplEnergy
    \lb\stress_{\elemIndex}\rb
    = 
    \sum_{\elemIndex=1}^{\nElem} \complStrainEnergy_\elemIndex\lb\stress_{\elemIndex}\rb
    +
    \sum_{\elemIndex=1}^{\nElem} \complExternal_\elemIndex
    .
\end{equation}
\par
As has been described in \Cref{subsec:principleMinimumComplementaryEnergy}, the minimization of $\totalComplEnergy$ is constrained through the set of statically admissible stress fields $\setStaticallyAdmissible$. In the same manner, we define the set of statically admissible stress fields per element $\setStaticallyAdmissible_{\elemIndex}$ such that the equilibrium conditions
\begin{equation}
    \frac{\partial\stress_{\elemIndex}(\x)}{\partial \x} 
    + \bodyForceDensity
    = 0,
    \quad
    \text{in } \left[x_{\elemIndex},x_{\elemIndex+1}\right],
    \quad 
    \elemIndex=1,\dots,\nElem
    ,
    \label{eq:equilibriumElement}
\end{equation}
and the traction boundary conditions from \Cref{eq:couplingConditions,eq:noTractionBC} are fulfilled.
Analogously to \Cref{eq:structuralAnalysisProblem}, we state the structural analysis problem for the composed rod in minimization form as
\begin{equation}
    \min_{\stress_{\elemIndex}\in\setStaticallyAdmissible_{\elemIndex}} \totalComplEnergy\lb\stress_{\elemIndex}\rb.
    \label{eq:structuralAnalysisProblemElem}
\end{equation}
\par
For the structural design optimization problem, we define the element-specific variants of the design variables and the set of admissible designs as $\designVarVector_{\elemIndex}$ and $\setAdmissibleDesign_{\elemIndex}$, respectively. In the case of the rod considered here, the design variables could be given, e.g., by Young's moduli $\youngModulus[\elemIndex]$ or cross-sectional areas $\crossSectionalArea[\elemIndex]$. As mentioned above, we will consider the latter in this paper. Regardless of the specific choice for the design variables, the structural design optimization problem can be stated as
\begin{equation}
    \min_{\designVarVector_{\elemIndex}\in\setAdmissibleDesign_{\elemIndex}}
    \min_{\stress_{\elemIndex}\in\setStaticallyAdmissible_{\elemIndex}}\totalComplEnergy\lb\stress_{\elemIndex};\designVarVector_{\elemIndex}\rb.
    \label{eq:designOptimizationProblemElem}
\end{equation}

\subsection{\gls{qubo} Formulations for a Rod under Self-Weight Loading}
\label{subsec:qubo1DRod}
Based on the problem formulations of the structural analysis and design optimization problem for the rod in minimization form given in \Cref{eq:structuralAnalysisProblemElem,eq:designOptimizationProblemElem}, we can derive the corresponding problem statements in \gls{qubo} form. In this regard, we already keep in mind the limitations of current \gls{qa} hardware and, e.g., try to keep the number of required qubits as low as possible. 
\par
For the problem presented in \Cref{subsec:problemFormulation1DRod}, it turns out to be advantageous to switch from a description in terms of stresses $\stress_{\elemIndex}(\x)$ to an equivalent description in terms of undirected force functions $\force_{\elemIndex}(\x)$, i.e.,
\begin{equation}
\force_{\elemIndex}(\x) 
=
\lvert\stress_{\elemIndex}(\x) \cdot \normal \rvert
\crossSectionalArea[\elemIndex],
\label{eq:force}
\end{equation}
with unit normal vector $\normal \in \{-1,1\}$.
Before we can explain the effects of this step, we first need to introduce the finite-dimensional approximation of $\force_{\elemIndex}(\x)$ through linear interpolation functions $\basisFct^{\mathrm{I/II}}_{\elemIndex}(\x)$ and scalar, real-valued coefficients $\basisCoeff^{\mathrm{I/II}}_{\elemIndex}\in\mathbb{R}$:
\begin{equation}
    \force_{\elemIndex}\lp\x\rp
    \approx
    \basisCoeff^{\mathrm{I}}_\elemIndex \basisFct^{\mathrm{I}}_\elemIndex(\x)
    +
    \basisCoeff^{\mathrm{II}}_{\elemIndex} \basisFct^{\mathrm{II}}_{\elemIndex}(\x),
    \label{eq:ansatzForce}
\end{equation}
with
\begin{equation}
    \basisFct^{\mathrm{I}}_{\elemIndex}(\x) = \frac{\x_{\elemIndex+1}-\x}{\x_{\elemIndex+1}-\x_{\elemIndex}},
    \quad
    \basisFct^{\mathrm{II}}_{\elemIndex}(\x) = \frac{x-\x_{\elemIndex}}{\x_{\elemIndex+1}-\x_{\elemIndex}}.
    \label{eq:basisFunctions}
\end{equation}
Since a \gls{qubo} problem needs to be defined in binary variables, we represent each real-valued coefficient $\basisCoeff^{\mathrm{I/II}}_{\elemIndex}$ using a vector of qubits $\qubitsVectorCoeffs_{\elemIndex}$ that collects binary variables $\qubitCoeff_l\in\{0,1\}$. We restrict ourselves to cases in which it holds that
$\basisCoeff^{\mathrm{I}/\mathrm{II}}_{\elemIndex}\in[0,1]$ and use the following representation using $\nQubitsPerNode$ qubits:
\begin{equation}
    \basisCoeff^{\mathrm{I}/\mathrm{II}}_{\elemIndex}\lb\qubitsVectorCoeffs_{\elemIndex}\rb
    =
    \frac{1}{2^{\nQubitsPerNode}-1}
    \sum_{l=1}^{\nQubitsPerNode}
     2^l \qubitCoeff_l 
     .
     \label{eq:representationBinary}
\end{equation}
\par
Note that, due to the form of the interpolation functions, the coefficients $\basisCoeff^{\mathrm{I}}_\elemIndex$ and $\basisCoeff^{\mathrm{II}}_\elemIndex$ correspond to the nodes $\nodeIndex$ and  $\nodeIndex+1$, respectively.
As a result, formulating the problem in terms of forces has the following advantage. At the interface, i.e., node, between two elements $\elemIndex-1$ and $\elemIndex$, we can merge the corresponding two coefficients into one nodal coefficient: 
\begin{equation}
   \basisCoeff^{\mathrm{II}}_{\elemIndex-1}\lb\qubitsVectorCoeffs_{\elemIndex-1}\rb 
   = 
   \basisCoeff^{\mathrm{I}}_{\elemIndex}\lb\qubitsVectorCoeffs_{\elemIndex}\rb 
   = 
   \basisCoeff_{\nodeIndex}\lb\qubitsVectorCoeffs_{\nodeIndex}\rb, 
   \label{eq:nodalBasisCoeff}
\end{equation}
where $\basisCoeff_{\nodeIndex}\lb\qubitsVectorCoeffs_{\nodeIndex}\rb$ are the nodal coefficients located at coordinates $\x_{\nodeIndex}$ using the same binary representation as stated in \Cref{eq:representationBinary} and a corresponding vector of qubits $\qubitsVectorCoeffs_{\nodeIndex}$ with $\qubitCoeff_{i}\in\{0,1\}$. Consequently, the force equilibrium from \Cref{eq:equilibriumForces} is satisfied by design, and the number of unknown coefficients, and thus the number of qubits, can be significantly reduced, by a factor of about two.
\bigskip\par 
So, using \Cref{eq:internalComplStrainEnergyElement,eq:force,eq:ansatzForce,eq:basisFunctions,eq:representationBinary,eq:nodalBasisCoeff}, we can first derive an expression for the internal complementary energy on an element $\elemIndex$ with nodes $\nodeIndex$ and $\nodeIndex+1$ in terms of the corresponding vectors of qubits $\qubitsVectorCoeffs_{\nodeIndex}$ and $\qubitsVectorCoeffs_{\nodeIndex+1}$:
\begin{equation}
    \complStrainEnergy_{\elemIndex}
    \lb
        \qubitsVectorCoeffs_{\nodeIndex},
        \qubitsVectorCoeffs_{\nodeIndex+1} 
    \rb
    =
    \frac{1}{2}
    \intElement{\elemIndex}
    {
        \frac{1}{\youngModulus[\elemIndex]\crossSectionalArea[\elemIndex]}\force_{\elemIndex}^2\lb\qubitsVectorCoeffs_{\nodeIndex},\qubitsVectorCoeffs_{\nodeIndex+1} \rb(\x)
    }
    \label{eq:complStrainEnergyQubits}
\end{equation}
What remains is to incorporate the conditions for statically admissible stress fields (cf. \Cref{eq:equilibriumElement,eq:noTractionBC}), re-formulated in terms of forces as well. 
The zero-traction boundary condition from \Cref{eq:noTractionBC} is directly considered by setting
\begin{equation}
    \basisCoeff_{\nElem+1} = 0,
\end{equation}
without using any qubits for this coefficient.
The equilibrium constraints stated in \Cref{eq:equilibriumElement} are taken into account by introducing a \textit{penalty term} $\penaltyTerm$. Starting from \Cref{eq:equilibriumElement} and using \Cref{eq:force,eq:ansatzForce,eq:basisFunctions}, we can derive the element-wise contributions
\begin{equation}
    \penaltyTerm_{\elemIndex}
    \lb
        \qubitsVectorCoeffs_{\nodeIndex},
        \qubitsVectorCoeffs_{\nodeIndex+1} 
    \rb
    =
    \frac{
            \basisCoeff_{\nodeIndex+1}
            \lb
                \qubitsVectorCoeffs_{\nodeIndex+1}
            \rb
            -
            \basisCoeff_{\nodeIndex}
            \lb
                \qubitsVectorCoeffs_{\nodeIndex}
            \rb
        }
        {
            \crossSectionalArea[\elemIndex]
        } 
    + \lp\x_{\nodeIndex+1}-\x_{\nodeIndex}\rp\bodyForceDensity
    \overset{!}{=}
    0,
    \quad 
    \elemIndex=1,\dots,\nElem
    .
    \label{eq:penaltyTermElement}
\end{equation}
By collecting all vectors of qubits for the nodal coefficients $\qubitsVectorCoeffs_{\nodeIndex}$ for $\nodeIndex=1,\dots,\nElem+1$ into a vector of qubits $\qubitsVectorCoeffs$, we can combine the contributions into one penalty term:
\begin{equation}
    \penaltyTerm\lb\qubitsVectorCoeffs\rb 
    = 
    \frac{1}{\nElem}
    \sum_{\elemIndex=1}^{\nElem} \lp\penaltyTerm_{\elemIndex}\lb\qubitsVectorCoeffs_{\nodeIndex},
        \qubitsVectorCoeffs_{\nodeIndex+1} \rb\rp^2. 
    \label{eq:penaltyTermQubits}
\end{equation}

\par
In addition, we introduce a vector $\qubitsVector$ that contains all the qubit variables present in a problem, where for the structural analysis problem we have $\qubitsVector=\qubitsVectorCoeffs$. Consequently, we define the objective function $\objective\lb\qubitsVector\rb$ that accounts both for the minimization of the total complementary energy $\totalComplEnergy\lb\qubitsVector\rb$ and the constraints from $\setStaticallyAdmissible_{\elemIndex}$ through the penalty term $\penaltyTerm\lb\qubitsVector\rb$:
\begin{equation}
    \objective\lb\qubitsVector\rb = \totalComplEnergy\lb\qubitsVector\rb + \penaltyWeight \penaltyTerm\lb\qubitsVector\rb,
    \label{eq:objectiveQubits}
\end{equation}
where $\penaltyWeight$ is the \textit{penalty weight}.
Finally, the \textit{structural analysis problem in \gls{qubo} form} is given as
\begin{equation}
    \min_{\qubitsVector} \objective\lb\qubitsVector\rb.
    \label{eq:structuralAnalysisProblemQubits}
\end{equation}
This problem is in \gls{qubo} form, since (i) the objective function $\objective\lb\qubitsVector\rb$ only contains combinations of qubits that are quadratic (cf. \Cref{eq:complStrainEnergyQubits,eq:penaltyTermQubits}), (ii) all constraints originally given are incorporated via the penalty term in \Cref{eq:objectiveQubits}, (iii) all remaining unknowns are described by qubits and, thus, binary variables, and, (iv) \Cref{eq:structuralAnalysisProblemQubits} is an optimization problem.
\bigskip\par
Based on the formulation of the structural analysis problem, we will now derive the structural design optimization problem. In particular, we consider the cross-sectional areas of each element $\crossSectionalArea[\elemIndex]$ as the design variables $\designVarVector_{\elemIndex}$ and allow them to be chosen from a set of admissible designs, here a set of two choices:
\begin{equation}
    \setAdmissibleDesign=\{\crossSectionalAreaChoice{1}, \crossSectionalAreaChoice{2}\},
\end{equation}
where $\crossSectionalAreaChoice{1}$ and $\crossSectionalAreaChoice{2}$ are two options for each cross-sectional area $\crossSectionalArea[\elemIndex]$. The representation of this quantity through qubits is discussed next.
As can be seen in \Cref{eq:complStrainEnergyQubits,eq:penaltyTermElement}, the cross-sectional area appears only in the denominator. 
Therefore, we use the following representation in terms of a single qubit variable $\qubitDesign_{\elemIndex}\in\{0,1\}$ to choose $\crossSectionalArea[\elemIndex]$ between $\crossSectionalAreaChoice{1}$ and $\crossSectionalAreaChoice{2}$ for each element:
\begin{equation}
    \frac{1}{\crossSectionalArea[\elemIndex]}
    = 
    \frac{1}{\crossSectionalAreaChoice{1}} 
    +
    \lp
        \frac{1}{\crossSectionalAreaChoice{2}}-\frac{1}{\crossSectionalAreaChoice{1}}
    \rp\qubitDesign_e.
    \label{eq:representationDesign}
\end{equation}
Similar to above, we collect all the qubits $\qubitDesign_e$ for $\elemIndex=1,\dots,\nElem$ in a corresponding vector of qubits $\qubitsVectorDesigns$.
\par
Now that additional variables $\crossSectionalArea[\elemIndex]$ are introduced in terms of qubits $\qubitDesign_{\elemIndex}$, the expressions for the complementary strain energy from \Cref{eq:complStrainEnergyQubits} and the penalty terms from \Cref{eq:penaltyTermElement} contain combinations of qubits that are up to order three. Thus, the same holds for the objective function $\objective\lb\qubitsVectorCoeffs,\qubitsVectorDesigns\rb$, now depending on both $\qubitsVectorCoeffs$ and $\qubitsVectorDesigns$. To recover a quadratic form, we make use of so-called \textit{quadratization} or \textit{degree reduction techniques}~\cite{Dattani2019}. As a drawback of this step, a number of auxiliary qubit variables $\qubitsVectorAuxiliary$ needs to be introduced~\cite{Yarkoni2022}. Applying one of these techniques to the objective function $\objective\lb\qubitsVectorCoeffs,\qubitsVectorDesigns\rb$ will yield a quadratic version $\objectiveQuadratic\lb\qubitsVectorCoeffs,\qubitsVectorDesigns,\qubitsVectorAuxiliary\rb$, which in addition depends on the auxiliary qubits $\qubitsVectorAuxiliary$. Finally, the \textit{structural design optimization problem in \gls{qubo} form} reads:
\begin{equation}
    \min_{\qubitsVectorCoeffs,\,\qubitsVectorDesigns,\,\qubitsVectorAuxiliary} \objectiveQuadratic\lb\qubitsVectorCoeffs,\qubitsVectorDesigns,\qubitsVectorAuxiliary\rb.
\end{equation}
In the following, we will also refer to the qubits that originally result from the problem formulation, i.e., $\qubitsVectorCoeffs$ and $\qubitsVectorDesigns$, as \textit{input qubits}, while using the term \textit{logical qubits} for all the qubits present in the final problem formulation $\qubitsVector$, potentially including the auxiliary qubits $\qubitsVectorAuxiliary$ from the degree reduction.  We will denote the number of corresponding input or logical qubits by $\nQubitsInput$ and $\nQubitsLogical$, respectively.

\section{Results}
\label{sec:results}
In order to demonstrate the suitability of the presented formulation for solving both structural analysis and design optimization problems using \gls{qa}, we present results for two test cases. In the first test case (\Cref{subsec:resultsStructuralAnalysis}), we consider the structural analysis problem for the rod composed of multiple elements, in this case still with fixed cross-sectional areas. In the second case (\Cref{subsec:resultsDesignOptimization}), we allow that each cross-sectional area can be chosen independently to find the optimal design of minimum compliance. For both cases, we discuss the features of the resulting \gls{qubo} formulations and assess the quality of the results by means of the respective analytic solutions.
\par
All of the results have been calculated using the open source software \textit{EngiOptiQA}~\cite{engioptiqa2023}, which we have developed in the course of this work. 
For the formulation of the \gls{qubo} model, the software relies on the \textit{Fixstars Amplify} \gls{sdk}~\cite{FixstarsAmplify2023}, whereas the solution process is performed via \textit{D-Wave's Ocean} \gls{sdk}~\cite{DWaveOcean2023}.
\subsection{Results for the Structural Analysis Problem}
\label{subsec:resultsStructuralAnalysis}
\begin{figure}
     \centering
     \begin{subfigure}[b]{0.49\textwidth}
        \centering
            \begin{tikzpicture}[scale=0.8]
                \fill[pattern=north east lines] (0,0) rectangle (3,0.5);
                \draw[thick] (0,0) -- (3,0);
                \draw[-latex] (0.25,0) -- (0.25, -0.75) node [right] {$\x$};
                \draw[latex-latex] (0.75,0) -- node[left]{\rodLength}(0.75,-5);
                \draw (1,0) -- (1,-5) -- (2,-5) -- (2,0);
                \foreach \y in {-1,-2,-3,-4} {
                    \draw (1,\y) -- (2,\y);
                }
                \foreach \y in {-1,-2,-3,-4,-5} {
                    \draw[-latex] (1.5, \y+0.9) -- (1.5, \y+0.1);
                }
                \draw[-latex] (2.5, -1.5) -- node [right] {$\bodyForceDensity$}(2.5, -2.3) ;
        
                \draw  (1.75,-5.5) node [below right] {$\crossSectionalArea$} -- (1.5,-5.1);
          
            \end{tikzpicture}
        \caption{Setup for the composed rod with identical cross sections.}
        \label{fig:structuralAnalysisSetup}
     \end{subfigure}
     \hfill
     \begin{subfigure}[b]{0.49\textwidth}
         \centering
\begin{tikzpicture}[scale=0.75]

\definecolor{darkgray176}{RGB}{176,176,176}

\begin{axis}[
title={\acrshort{qubo} Pattern},
tick align=outside,
tick pos=left,
xlabel = Index $j$,
xmin=-0.5, xmax=49.5,
xtick style={color=black},
ylabel = Index $i$,
y dir=reverse,
ymin=-0.5, ymax=49.5,
ytick style={color=black}
]
\addplot graphics [includegraphics cmd=\pgfimage,xmin=-0.5, xmax=49.5, ymin=49.5, ymax=-0.5] {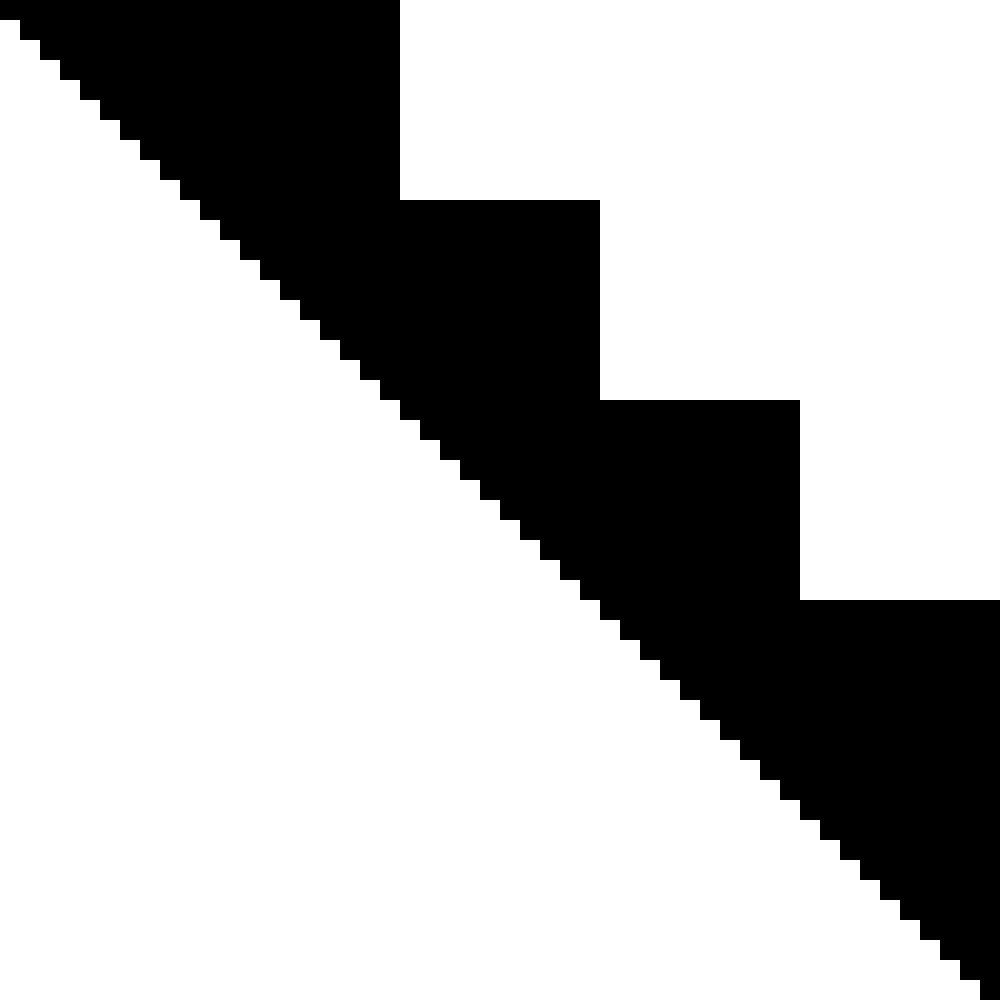};

\draw[thick, decorate,decoration={brace,amplitude=5pt,mirror}]
    (axis cs:-0.5,20) -- (axis cs:19.5,20) node[midway, yshift=-12pt] {$2\nQubitsPerNode$};
\draw[thick, decorate,decoration={brace,amplitude=5pt}]
    (axis cs:30,-0.5) -- (axis cs:30,19.5) node[midway, xshift=15pt] {$2\nQubitsPerNode$};

\draw[thick, decorate,decoration={brace,amplitude=3pt}]
    (axis cs:19.5,9.5) -- (axis cs:29.5,9.5);
\draw[thick, decorate,decoration={brace,amplitude=3pt}]
    (axis cs:20,-0.5) -- (axis cs:20,9) node[midway, xshift=18pt, yshift=-3pt] {$\nQubitsPerNode$};
    
\draw[purple, line width= 3pt] (axis cs:-0.5,-0.5) -- (axis cs:19.5,-0.5) -- (axis cs:19.5,19.5) -- cycle;
\draw[lightgray, line width= 2pt, loosely dashed] (axis cs:9.5,9.5) -- (axis cs:29.5,9.5) -- (axis cs:29.5,29.5) -- cycle;
\draw[lightgray, line width= 2pt, loosely dashed]  (axis cs:19.5,19.5) -- (axis cs:39.5,19.5) -- (axis cs:39.5,39.5) -- cycle;
\draw[lightgray, line width= 2pt, loosely dashed]  (axis cs:29.5,29.5) -- (axis cs:49.5,29.5) -- (axis cs:49.5,49.5) -- cycle;

\end{axis}
\end{tikzpicture}
         \caption{Pattern of the interactions between the input qubits $\qubitSymbol_i$ and $\qubitSymbol_j$.}
         \label{fig:structuralAnalysisQUBOPattern}
     \end{subfigure}
        \caption{\centering Structural Analysis Problem: Setup and \gls{qubo} pattern.}
        \label{fig:structuralAnalysisProblemSetupAndQUBOPattern}
\end{figure}
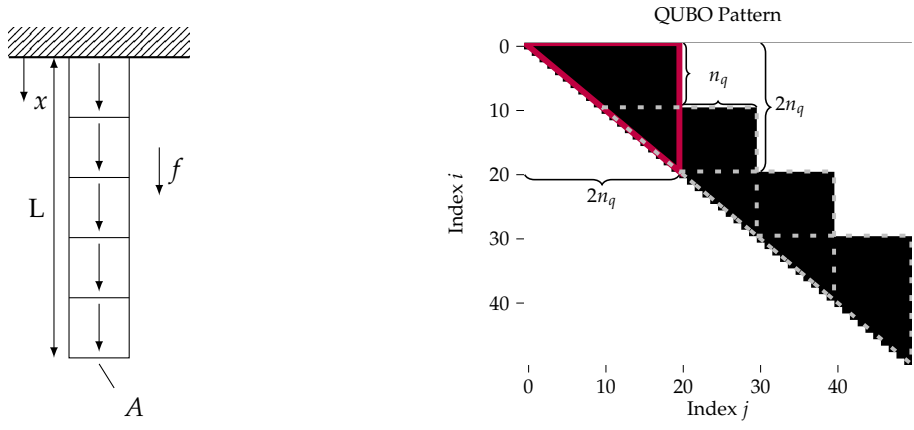

For the structural analysis problem, we consider a rod of length $\rodLength = 1.5$, which is composed of $\nElem=5$ elements with identical cross-sectional areas $\crossSectionalArea[\elemIndex] \equiv\crossSectionalArea=0.25$ and Young's moduli $\youngModulus[\elemIndex]\equiv\youngModulus=1.0$. 
The rod is subject to self-weight loading through the body force density $\bodyForceDensity = 1.5$. 
The values of all relevant quantities are given in \Cref{tab:structuralAnalysisProblem} and a sketch of the setup of the problem is provided in \Cref{fig:structuralAnalysisSetup}.
In this case, the analytic solution for the force function $ F^*(\x)$ can be found to be 
\begin{equation}
    F^*(\x) = \crossSectionalArea \bodyForceDensity (\rodLength - \x).
\end{equation}
\begin{table} 
\caption{Structural Analysis Problem: test case settings.\label{tab:structuralAnalysisProblem}}
\newcolumntype{C}{>{\centering\arraybackslash}X}
\begin{tabularx}{\textwidth}{CCCCC}
\toprule
$\rodLength$	& $\nElem$	& $\crossSectionalArea$ & $\youngModulus$ & $\bodyForceDensity$ \\
\midrule
1.5	& 5			& 0.25 & 1.0 & 2.5 \\
\bottomrule
\end{tabularx}
\end{table}
\par
Next, we will consider the \gls{qubo} form of the problem, which has been derived in \Cref{subsec:problemFormulation1DRod}.
For the binary representation of $\basisCoeff_{\nodeIndex}\lb\qubitsVectorCoeffs_{\nodeIndex}\rb$ given in \Cref{eq:representationBinary}, we use $\nQubitsPerNode=10$ qubits, which sums up to $\nQubitsLogical=50$ logical qubits in $\qubitsVector$ for the entire problem. The corresponding pattern is shown in \Cref{fig:structuralAnalysisQUBOPattern} and renders a diagonal matrix of overlapping blocks, where each triangular block, such as the one outlined in red, represents the interaction between two adjacent nodes on an element. Consequently, the block width is always $2\nQubitsPerNode$. This local character of interactions results in the visible sparsity pattern and will also be maintained when scaling up the problem, e.g., by increasing the number of rod elements. 
\bigskip\par
After specifying the \gls{qubo} problem for the structural analysis problem, we now discuss relevant aspects of solving it on \gls{qa} hardware and present a strategy that we developed for this purpose.
For solving the problem on \gls{qa} hardware, we have used the \textit{Leap™} quantum cloud service offered by \textit{D-Wave}~\cite{DWaveLeap2023}.
In particular, computations have been performed on the \texttt{Advantage\_system4.1} machine, which uses the \textit{Advantage performance update} \gls{qpu}. The latter is based on the so-called \textit{Pegasus} architecture~\cite{Boothby2020}, which consists of a lattice of \textit{physical qubits} and connections between pairs of them. Due to the probabilistic nature of \gls{qa}, the solution process involves running the annealing phase multiple times. The corresponding number of \textit{samples}, or \textit{reads}, will be referred to as $\nreads$, with the \textit{annealing time} per read denoted by $\tAnnealing$. 
\bigskip\par
On the hardware level, the probability of an individual physical qubit being in a particular state can be influenced by \textit{biases}, while the preference for certain combinations of binary values is shaped by \textit{couplers} that yield a corresponding interaction strength.
The values for these biases and coupling strengths are given by the linear and quadratic coefficients of the \gls{qubo} problem. However, there may not be a one-to-one correspondence between the \gls{qpu} and the variables and interactions in the \gls{qubo} problem. 
This is the reason why a \textit{minor-embedding}, i.e., a mapping of the logical qubits and interactions of the \gls{qubo} problem onto the physical qubits and couplers of the \gls{qpu}, is required. For example, it may be necessary to represent a single logical qubit by a group of physical qubits forming a so-called \textit{chain}. Consequently, all physical qubits in such a chain need to have the same value for a consistent sample of the logical variable. To achieve this, the qubits in the chain are coupled more strongly than for interactions with other qubits, increasing the probability that all qubits in the chain will have identical values.
Note, however, that we do not have to take care of the minor-embedding manually but it is automated through the \textit{Ocean} software provided by \textit{D-Wave}.
\bigskip\par
Another practically relevant aspect is the fact that the \gls{qpu} only supports a certain range of values for the biases and coupling strengths, i.e., the linear and quadratic coefficients. Consequently, only finite precision is provided in these coefficients. In this regard, the real-valued and high-precision coefficients resulting from the proposed formulation pose a challenge and give great importance to the scaling of the problem. 
The latter is especially influenced by the penalty weight $\penaltyWeight$.
While one would like to set this weight as high as possible to ensure fulfillment of the constraints, i.e., static admissibility, this strategy may not be useful on \gls{qa} hardware. The reason for this is that this approach would lead to large differences in the magnitudes of the penalty term $\penaltyWeight\penaltyTerm$ and the complementary energy $\totalComplEnergy$, which are present in the objective function $\objective$. As a consequence of the imprecision issue mentioned above, a proper minimization of $\totalComplEnergy$ is no longer ensured in this case. 
\par
To address this problem, we developed the following strategy. First, we relax the constraint for static admissibility and choose a ``small'' penalty weight $\penaltyWeight = \penaltyWeight_{\mathrm{small}}$ to perform \gls{qa} for a number of reads $\nreads$. This yields a set of solutions with near minimum complementary energy, which may slightly violate static admissibility. To filter out solutions violating the constraint, we apply an additional post-processing step on classical hardware. In particular, we update the problem formulation with an increased penalty weight $\penaltyWeight = \penaltyWeight_{\mathrm{large}}$. For the updated problem, we use each \gls{qa} sample as an initial point for a fast local search to identify the nearest minimum. This search is based on a steepest descent algorithm, where at each step the direction to descend is determined by the variable flip that induces the most significant reduction in the objective function. From the resulting set of improved samples, the optimal solution, i.e., the solution that has the minimum complementary energy and is statically admissible, can then be selected as the sample with the minimum value of the objective function $\objective$. Given the logical qubits from this sample, we can evaluate the ansatz for the force function $\force(\x)$ to obtain the solution to the structural analysis problem. 
\bigskip\par
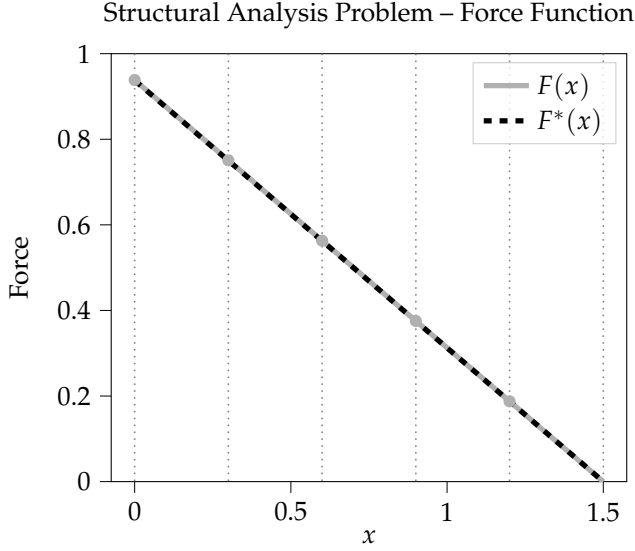
\begin{figure}
    \centering
\begin{tikzpicture}

\definecolor{darkgray176}{RGB}{176,176,176}
\definecolor{darkviolet1910191}{RGB}{176,176,176}
\definecolor{gray}{RGB}{128,128,128}
\definecolor{lightgray204}{RGB}{204,204,204}

\begin{axis}[
legend cell align={left},
legend style={fill opacity=0.8, draw opacity=1, text opacity=1, draw=lightgray204},
tick align=outside,
tick pos=left,
title={Structural Analysis Problem -- Force Function},
x grid style={darkgray176},
xlabel={$\x$},
xmin=-0.075, xmax=1.575,
xtick={0,0.5,1.0,1.5},
xtick style={color=black},
y grid style={darkgray176},
ylabel={Force},
ymin=0, ymax=1,
ytick style={color=black}
]
\addplot [semithick, gray, dotted, forget plot]
table {%
0 -0.0469208211143695
0 1.0
};
\addplot [semithick, gray, dotted, forget plot]
table {%
0.3 -0.0469208211143695
0.3 1.0
};
\addplot [semithick, gray, dotted, forget plot]
table {%
0.6 -0.0469208211143695
0.6 1.0
};
\addplot [semithick, gray, dotted, forget plot]
table {%
0.9 -0.0469208211143695
0.9 1.0
};
\addplot [semithick, gray, dotted, forget plot]
table {%
1.2 -0.0469208211143695
1.2 1.0
};
\addplot [semithick, gray, dotted, forget plot]
table {%
1.5 -0.0469208211143695
1.5 1.0
};
\addplot [line width=2pt, darkviolet1910191]
table {%
0 0.93841642228739
0.0333333333333333 0.917562724014337
0.0666666666666667 0.896709025741284
0.1 0.875855327468231
0.133333333333333 0.855001629195178
0.166666666666667 0.834147930922124
0.2 0.813294232649071
0.233333333333333 0.792440534376018
0.266666666666667 0.771586836102965
0.3 0.750733137829912
0.3 0.750733137829912
0.333333333333333 0.729879439556859
0.366666666666667 0.709025741283806
0.4 0.688172043010753
0.433333333333333 0.6673183447377
0.466666666666667 0.646464646464646
0.5 0.625610948191593
0.533333333333333 0.60475724991854
0.566666666666667 0.583903551645487
0.6 0.563049853372434
0.6 0.563049853372434
0.633333333333333 0.542196155099381
0.666666666666667 0.521342456826328
0.7 0.500488758553275
0.733333333333333 0.479635060280222
0.766666666666667 0.458781362007168
0.8 0.437927663734115
0.833333333333333 0.417073965461062
0.866666666666667 0.396220267188009
0.9 0.375366568914956
0.9 0.375366568914956
0.933333333333333 0.354512870641903
0.966666666666667 0.33365917236885
1 0.312805474095797
1.03333333333333 0.291951775822744
1.06666666666667 0.27109807754969
1.1 0.250244379276637
1.13333333333333 0.229390681003584
1.16666666666667 0.208536982730531
1.2 0.187683284457478
1.2 0.187683284457478
1.23333333333333 0.166829586184425
1.26666666666667 0.145975887911372
1.3 0.125122189638319
1.33333333333333 0.104268491365266
1.36666666666667 0.0834147930922127
1.4 0.0625610948191596
1.43333333333333 0.0417073965461064
1.46666666666667 0.0208536982730533
1.5 2.22044604925031e-16
};
\addlegendentry{$\force(\x)$}

\addplot [semithick, darkviolet1910191, mark=*, mark size=2, mark options={solid}, only marks, forget plot]
table {%
0 0.93841642228739
};
\addplot [semithick, darkviolet1910191, mark=*, mark size=2, mark options={solid}, only marks, forget plot]
table {%
0.3 0.750733137829912
};
\addplot [semithick, darkviolet1910191, mark=*, mark size=2, mark options={solid}, only marks, forget plot]
table {%
0.6 0.563049853372434
};
\addplot [semithick, darkviolet1910191, mark=*, mark size=2, mark options={solid}, only marks, forget plot]
table {%
0.9 0.375366568914956
};
\addplot [semithick, darkviolet1910191, mark=*, mark size=2, mark options={solid}, only marks, forget plot]
table {%
1.2 0.187683284457478
};
\addplot [line width=2pt, dashed, black]
table {%
0 0.9375
0.0333333333333333 0.916666666666667
0.0666666666666667 0.895833333333333
0.1 0.875
0.133333333333333 0.854166666666667
0.166666666666667 0.833333333333333
0.2 0.8125
0.233333333333333 0.791666666666667
0.266666666666667 0.770833333333333
0.3 0.75
0.3 0.75
0.333333333333333 0.729166666666667
0.366666666666667 0.708333333333333
0.4 0.6875
0.433333333333333 0.666666666666667
0.466666666666667 0.645833333333333
0.5 0.625
0.533333333333333 0.604166666666667
0.566666666666667 0.583333333333333
0.6 0.5625
0.6 0.5625
0.633333333333333 0.541666666666667
0.666666666666667 0.520833333333333
0.7 0.5
0.733333333333333 0.479166666666667
0.766666666666667 0.458333333333333
0.8 0.4375
0.833333333333333 0.416666666666667
0.866666666666667 0.395833333333333
0.9 0.375
0.9 0.375
0.933333333333333 0.354166666666667
0.966666666666667 0.333333333333333
1 0.3125
1.03333333333333 0.291666666666667
1.06666666666667 0.270833333333333
1.1 0.25
1.13333333333333 0.229166666666667
1.16666666666667 0.208333333333333
1.2 0.1875
1.2 0.1875
1.23333333333333 0.166666666666667
1.26666666666667 0.145833333333333
1.3 0.125
1.33333333333333 0.104166666666667
1.36666666666667 0.0833333333333333
1.4 0.0625
1.43333333333333 0.0416666666666666
1.46666666666667 0.0208333333333333
1.5 0
};

\addlegendentry{$\forceAnalytic(\x)$}

\end{axis}

\end{tikzpicture}
    \caption{Structural Analysis Problem: solution for the force function $\force(\x)$ obtained by \gls{qa} compared to the analytical solution $\force^*(\x)$.}
    \label{fig:structuralAnalysisQuantumAnnealing}
\end{figure}

\begin{table}
\caption{Structural Analysis Problem: settings for \gls{qa}.\label{tab:structuralAnalysisProblemQA}}
\newcolumntype{C}{>{\centering\arraybackslash}X}
\begin{tabularx}{\textwidth}{CCCCCCC}
\toprule
$\nQubitsPerNode$ & $\nQubitsInput\equiv\nQubitsLogical$ &$\penaltyWeight_{\mathrm{small}}$ & $\penaltyWeight_{\mathrm{large}}$ & $\nreads$ & $\tAnnealing$ \\
\midrule
10 & 50 & $2\cdot10^1$ & $10^9$ & 500 & 400 $\mu s$\\
\bottomrule
\end{tabularx}
\end{table}
To solve the structural analysis problem for the rod, we used the settings collected in \Cref{tab:structuralAnalysisProblemQA}. The result for $\force(\x)$ is shown in \Cref{fig:structuralAnalysisQuantumAnnealing} next to the analytical solution $\force^*(\x)$. Both functions are visually in very good agreement.
Additionally, we consider the relative error in the H1 norm
\begin{equation}
    \epsilon_{\mathrm{H^1}} 
    =
    \frac
    {
        \lVert \force(\x)-\force^*(\x) \rVert_{\mathrm{H^1}}
    }
    {
        \lVert \force^*(\x) \rVert_{\mathrm{H^1}}
    }
    .
\end{equation}
The choice of the H1 norm is motivated by the fact that it naturally corresponds to the objective function in our formulation. 
More precisely, the contributions from the function values themselves and the values for the derivative to the H1 norm correspond to the complementary energy and the penalty term for statically permissible solutions, respectively.
Consequently, minimizing the objective function also minimizes the error in the H1 norm.
For the presented test case, the relative error in the H1 norm is $\epsilon_{\mathrm{H^1}} = 9.78\cdot10^{-4}$. This demonstrates that the proposed formulation is capable of solving the structural analysis problem using \gls{qa}. 
As mentioned before, this is an essential building block for solving the combined design optimization problem that will be considered next.
\subsection{Results for the Structural Design Optimization Problem}
\label{subsec:resultsDesignOptimization}
\begin{figure}
    \centering
    \begin{tikzpicture}[scale=0.9]
        \fill[pattern=north east lines] (0,0) rectangle (3,0.5);
        \draw[thick] (0,0) -- (3,0);
        \draw[-latex] (0.25,0) -- (0.25, -0.75) node [right] {$\x$};
        \draw[latex-latex] (0.75,0) -- node[left]{\rodLength}(0.75,-4);
        \draw (1,0) rectangle (2,-2);
        \draw (1.25,-2) rectangle (1.75,-4);

        \foreach \y in {-1.5,-3.5} {
            \draw[-latex] (1.5, \y+0.9) -- (1.5, \y+0.1);
        }
        \draw[-latex] (2.5, -1.5) -- node [right] {$\bodyForceDensity$}(2.5, -2.3) ;

        \node at (1.5,-4.2) [below] {$\crossSectionalArea[\elemIndex]\in\setAdmissibleDesign$};
  
    \end{tikzpicture}
    \caption{Structural Design Optimization Problem: Setup for composed rod with variable cross sections.}
    \label{fig:designOptimizationSetup}
\end{figure}
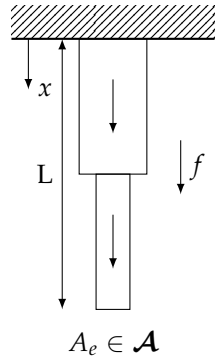

The structural design optimization problem is given through a size optimization of the compound rod introduced before. In particular, we choose the element-specific cross-sectional areas $\crossSectionalArea[\elemIndex]$ as design variables $\designVarVector_{\elemIndex}$. As described in \Cref{subsec:problemFormulation1DRod}, we have two choices $\crossSectionalAreaChoice{1}$ and $\crossSectionalAreaChoice{2}$ for the cross section of each component. 
All relevant quantities and a sketch of the setup of the problem are given in \Cref{tab:designOptimizationProblem} and  \Cref{fig:designOptimizationSetup}, respectively.
\begin{table}
\caption{Structural Design Optimization Problem: test case settings.\label{tab:designOptimizationProblem}}
\newcolumntype{C}{>{\centering\arraybackslash}X}
\begin{tabularx}{\textwidth}{CCCCC}
\toprule
$\rodLength$	& $\nElem$	& $\setAdmissibleDesign$ & $\youngModulus$ & $\bodyForceDensity$ \\
\midrule
1.5	& 2	& \{0.25,\,0.5\} & 1.0 & 2.5 \\
\bottomrule
\end{tabularx}
\end{table}
\bigskip
\par
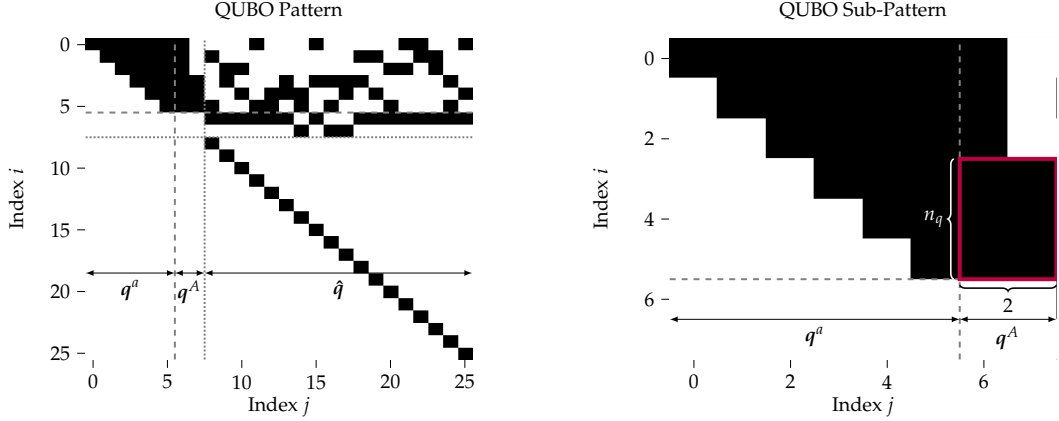
\begin{figure}
     \centering
     \begin{subfigure}[b]{0.49\textwidth}
         \centering
\begin{tikzpicture}[scale=0.75]

\definecolor{darkgray176}{RGB}{176,176,176}
\definecolor{gray}{RGB}{128,128,128}

\begin{axis}[
title={\acrshort{qubo} Pattern},
tick align=outside,
tick pos=left,
xlabel = Index $j$,
x grid style={darkgray176},
xmin=-0.5, xmax=25.5,
xtick style={color=black},
ylabel = Index $i$,
y dir=reverse,
y grid style={darkgray176},
ymin=-0.5, ymax=25.5,
ytick style={color=black}
]
\addplot graphics [includegraphics cmd=\pgfimage,xmin=-0.5, xmax=25.5, ymin=25.5, ymax=-0.5] {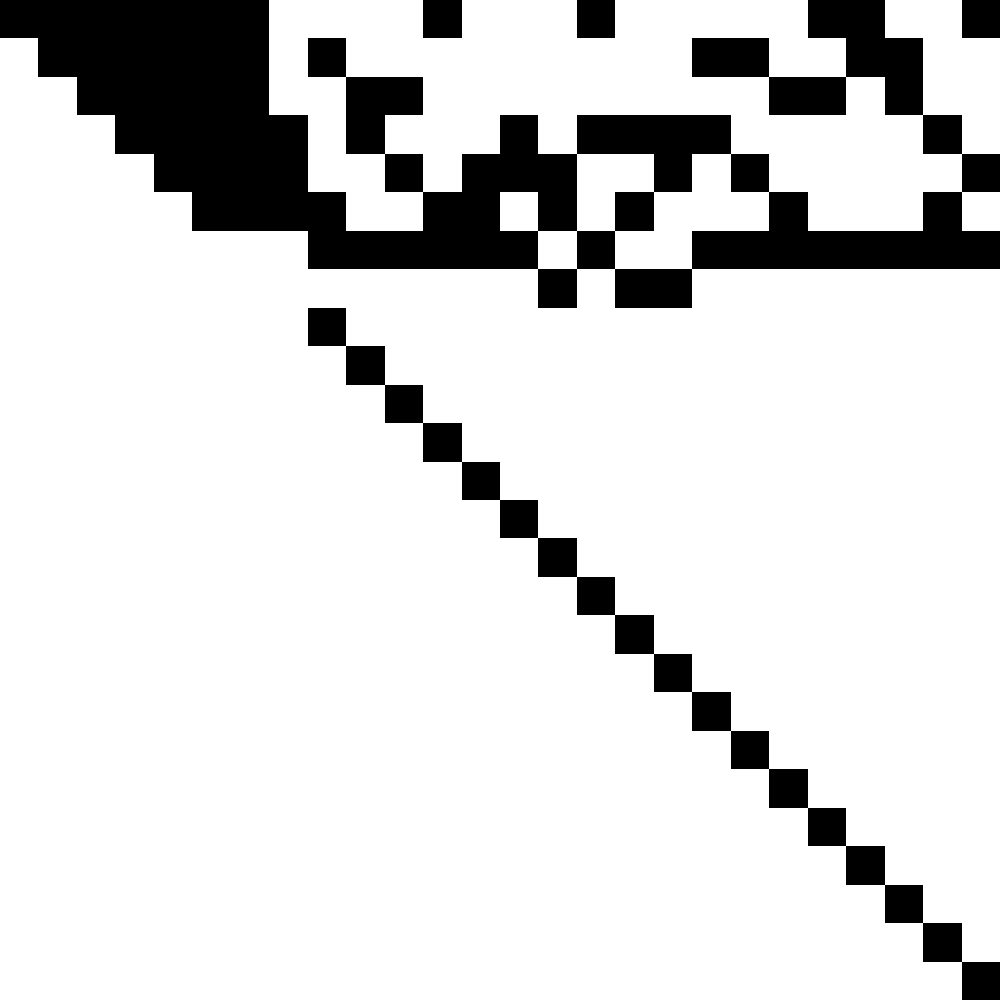};

\addplot [line width=1pt, gray, dashed]
table {%
5.5 25.5
5.5 -0.5
};
\addplot [line width=1pt, gray, dashed]
table {%
-0.5 5.5
25.5 5.5
};
\addplot [line width=1pt, gray, densely dotted]
table {%
7.5 25.5
7.5 -0.5
};
\addplot [line width=1pt, gray, densely dotted]
table {%
-0.5 7.5
25.5 7.5
};

\draw[latex-latex] (axis cs:-0.5,18.5) -- node [below] {$\qubitsVectorCoeffs$} (axis cs:5.5,18.5);
\draw[latex-latex] (axis cs:5.5,18.5) -- node [below] {$\qubitsVectorDesigns$} (axis cs:7.5,18.5);
\draw[latex-latex] (axis cs:7.5,18.5) -- node [below] {$\qubitsVectorAuxiliary$} (axis cs:25.5,18.5);
\end{axis}
\end{tikzpicture}
         \caption{Pattern of interactions between logical qubits, i.e., $\qubitsVectorCoeffs$, $\qubitsVectorDesigns$, and $\qubitsVectorAuxiliary$.}
         \label{fig:designOptimizationQUBOPattern}
     \end{subfigure}
     \hfill
     \begin{subfigure}[b]{0.49\textwidth}
         \centering
\begin{tikzpicture}[scale=0.75]

\definecolor{darkgray176}{RGB}{176,176,176}
\definecolor{gray}{RGB}{128,128,128}

\begin{axis}[
title={\acrshort{qubo} Sub-Pattern},
tick align=outside,
tick pos=left,
xlabel = Index $j$,
x grid style={darkgray176},
xmin=-0.5, xmax=7.5,
xtick style={color=black},
ylabel = Index $i$,
y dir=reverse,
y grid style={darkgray176},
ymin=-0.5, ymax=7.5,
ytick style={color=black}
]
\addplot graphics [includegraphics cmd=\pgfimage,xmin=-0.5, xmax=25.5, ymin=25.5, ymax=-0.5] {fig/design_optimization/design_optimization_problem_QUBO_pattern-000_hq.png};

\addplot [line width=1pt, gray, dashed]
table {%
5.5 7.5
5.5 -0.5
};

\addplot [line width=1pt,gray, dashed]
table {%
 7.5 5.5
-0.5 5.5 
};

\draw[line width =2pt, purple] (axis cs:5.5,2.5) rectangle (axis cs:7.5,5.5);
\draw[white,thick, decorate,decoration={brace,amplitude=3pt,mirror},xshift=-3pt]
    (axis cs:5.5,2.5) -- (axis cs:5.5,5.5) node[midway, xshift=-10pt, yshift=0pt] {$\nQubitsPerNode$};
\draw[thick, decorate,decoration={brace,amplitude=3pt,mirror}, yshift=-3pt]
    (axis cs:5.5,5.5) -- (axis cs:7.5,5.5) node[midway, xshift=0pt, yshift=-10pt] {$2$};
\draw[latex-latex] (axis cs:-0.5,6.5) -- node [below] {$\qubitsVectorCoeffs$} (axis cs:5.5,6.5);
\draw[latex-latex] (axis cs:5.5,6.5) -- node [below] {$\qubitsVectorDesigns$} (axis cs:7.5,6.5);
\end{axis}

\end{tikzpicture}
         \caption{Sub-pattern of interactions between input qubits, i.e., $\qubitsVectorCoeffs$ and $\qubitsVectorDesigns$.}
         \label{fig:designOptimizationQUBOSubPattern}
     \end{subfigure}
        \caption{\centering Structural Design optimization problem: \acrshort{qubo} patterns.}
        \label{fig:designOptimizationQUBOSubPatterns}
\end{figure}

For the \gls{qubo} problem, we use $\nQubitsPerNode=3$ qubits for the real-valued coefficients $\basisCoeff_{\nodeIndex}\lb\qubitsVectorCoeffs_{\nodeIndex}\rb$ (cf. \Cref{eq:representationBinary}) and one qubit $\qubitDesign_{\elemIndex}$ for each design variable $\crossSectionalArea[\elemIndex]$ (cf. \Cref{eq:representationDesign}), resulting in $\nQubitsInput=8$ input qubits.
As explained above, the resulting polynomial for the objective function $\objective$ involves terms up to order three in the input qubits $\qubitsVectorCoeffs$ and $\qubitsVectorDesigns$. To transform the cubic polynomial to a quadratic one, the \gls{ntr-kzfd}~\cite{Kolmogorov2004, Freedman2005} and the \gls{ptr} by Ishikawa~\cite{Ishikawa2011} are applied.
The degree reduction introduces a number of auxiliary qubits $\qubitsVectorAuxiliary$, leading to $\nQubitsLogical=26$ logical qubits, and corresponding additional interactions. This is reflected in the resulting pattern shown in \Cref{fig:designOptimizationQUBOPattern}. Ignoring the auxiliary qubits, the sub-pattern in \Cref{fig:designOptimizationQUBOSubPattern} on the one hand shows the block structure seen in the structural analysis problem and, on the other hand, a number of new blocks related to the qubits $\qubitDesign$ for the additional design variables for each element $\elemIndex$. These blocks have a maximum size of $\nQubitsPerNode\times 2$, since each element, i.e., each rod component, is formed by two adjacent nodes. Nevertheless, the characteristics of the final \gls{qubo} problem for the design optimization problem are different from those for the structural analysis problem discussed in the previous section due to the prerequisite application of a degree reduction technique.
\bigskip\par
We will discuss the consequences of this difference in terms of its practical impact on the use of \gls{qa} hardware in the following. The introduction of additional logical qubits and interactions by the degree reduction technique decreases the sparsity in the \gls{qubo} pattern. This makes the minor-embedding onto the \gls{qpu} more challenging. In fact, it may result in longer chains that are more susceptible to errors due to decoherence or thermal fluctuations, decreasing the quality of the \gls{qa} outcomes. 
\par
However, this quality is of special importance to the design optimization problem. The reason for this is the fact that the strategy of relaxing the penalty weight for the \gls{qa} and post-processing the results with a local search for an increased penalty weight cannot be used here. If we apply this strategy to the design optimization problem, suboptimal designs that satisfy the static admissibility constraint may be preferred to solutions with optimal designs but constraint violations during the post-processing step due to the large penalty weight. Yet, it is the optimal design that we are primarily interested in. Thus, we need to obtain high-quality outcomes directly from the \gls{qa} where the sample with minimum objective value represents the optimal design and fulfills the constraint with sufficient accuracy. 
\par
This requires that the susceptibility to inaccuracies or errors during the annealing phase be minimized by carefully setting up the \gls{qubo} problem. On the one hand, this limits the size of the penalty weight, since too large values will lead to precision problems, as explained above. On the other hand, we need to reduce the number of chains created in the course of the minor-embedding and their length as much as possible. This basically restricts the complexity of the \gls{qubo} problem that results from the additional qubits and interactions by the degree reduction technique.
Eventually, this limits the scale and the complexity of the problems that can be solved on the available \gls{qa} hardware and is the reason why we consider such a small-scale example here. Note, however, that this is not an inherent limitation of the formulation presented, and that future developments in hardware or problem formulation, as well as the use of other architectures, may address the above issues (see also \Cref{sec:discussion}).
\bigskip\par
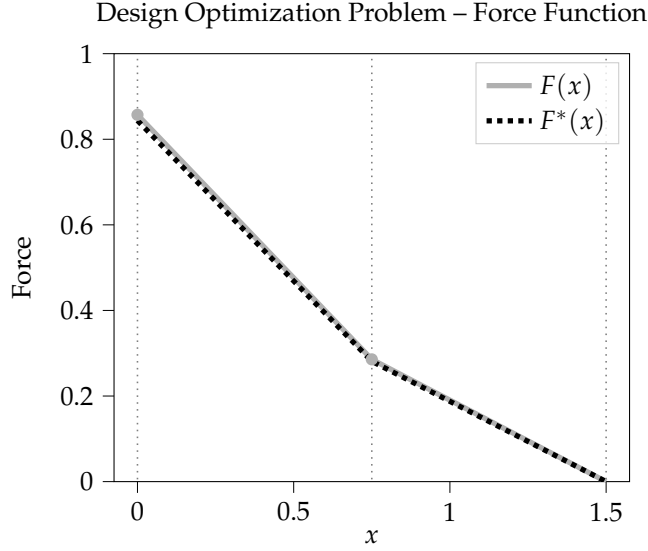
\begin{figure}
    \centering
\begin{tikzpicture}

\definecolor{darkgray176}{RGB}{176,176,176}
\definecolor{darkviolet1910191}{RGB}{176,176,176}
\definecolor{gray}{RGB}{128,128,128}
\definecolor{lightgray204}{RGB}{204,204,204}

\begin{axis}[
legend cell align={left},
legend style={fill opacity=0.8, draw opacity=1, text opacity=1, draw=lightgray204},
tick align=outside,
tick pos=left,
title={Design Optimization Problem -- Force Function},
x grid style={darkgray176},
xlabel={$\x$},
xmin=-0.075, xmax=1.575,
xtick={0,0.5,1.0,1.5},
xtick style={color=black},
y grid style={darkgray176},
ylabel={Force},
ymin=0, ymax=1,
ytick style={color=black}
]
\addplot [semithick, gray, dotted, forget plot]
table {%
0 -0.0428571428571429
0 1.0
};
\addplot [semithick, gray, dotted, forget plot]
table {%
0.75 -0.0428571428571429
0.75 1.0
};
\addplot [semithick, gray, dotted, forget plot]
table {%
1.5 -0.0428571428571429
1.5 1.0
};

\addplot [line width=2pt, darkviolet1910191]
table {%
0 0.857142857142857
0.0833333333333333 0.793650793650794
0.166666666666667 0.73015873015873
0.25 0.666666666666667
0.333333333333333 0.603174603174603
0.416666666666667 0.53968253968254
0.5 0.476190476190476
0.583333333333333 0.412698412698413
0.666666666666667 0.349206349206349
0.75 0.285714285714286
0.75 0.285714285714286
0.833333333333333 0.253968253968254
0.916666666666667 0.222222222222222
1 0.19047619047619
1.08333333333333 0.158730158730159
1.16666666666667 0.126984126984127
1.25 0.0952380952380952
1.33333333333333 0.0634920634920635
1.41666666666667 0.0317460317460319
1.5 0
};
\addlegendentry{$\force(\x)$}
\addplot [semithick, darkviolet1910191, mark=*, mark size=2, mark options={solid}, only marks, forget plot]
table {%
0 0.857142857142857
};
\addplot [semithick, darkviolet1910191, mark=*, mark size=2, mark options={solid}, only marks, forget plot]
table {%
0.75 0.285714285714286
};

\addplot [line width=2pt, dotted, black]
table {%
0 0.84375
0.0833333333333333 0.78125
0.166666666666667 0.71875
0.25 0.65625
0.333333333333333 0.59375
0.416666666666667 0.53125
0.5 0.46875
0.583333333333333 0.40625
0.666666666666667 0.34375
0.75 0.28125
0.75 0.28125
0.833333333333333 0.25
0.916666666666667 0.21875
1 0.1875
1.08333333333333 0.15625
1.16666666666667 0.125
1.25 0.09375
1.33333333333333 0.0625
1.41666666666667 0.03125
1.5 0
};
\addlegendentry{$\forceAnalytic(\x)$}
\end{axis}

\end{tikzpicture}
    \caption{Structural Design Optimization Problem: solution for the force function $\force(\x)$ obtained by \gls{qa} compared to the analytical solution $\force^*(\x)$.}
    \label{fig:designOptimizationQuantumAnnealing}
\end{figure}

Next, we present the results for the structural design optimization problem obtained using \gls{qa} with the settings from \Cref{tab:designOptimizationProblemQA}.
\begin{table}
\caption{Structural Design Optimization Problem: settings for \gls{qa}.\label{tab:designOptimizationProblemQA}}
\newcolumntype{C}{>{\centering\arraybackslash}X}
\begin{tabularx}{\textwidth}{CCCCCCCC}
\toprule
$\nQubitsPerNode$ &$\nQubitsInput$& $\nQubitsLogical$ &$\penaltyWeight$ & $\nreads$ & $\tAnnealing$\\
\midrule
3 & 8 & 26 & 5& 800 & 400 $\mu s$\\ 
\bottomrule
\end{tabularx}
\end{table}
Again, we will compare the solution $\force(\x)$ to the analytical solution $\force^*(\x)$. 
Here, the optimal design can be determined analytically as $\designVarVector_{\mathrm{opt}} = \lb\crossSectionalAreaChoice{2}, \crossSectionalAreaChoice{1}\rb=\lb0.5, 0.25\rb$. Using \gls{qa}, we also find this optimal design through the sample with the minimum value for the objective function. In addition, the corresponding force function is shown in \Cref{fig:designOptimizationQuantumAnnealing} and there is a good match between the numerical and analytical solutions. The respective relative error in the H1 norm amounts to $\epsilon_{\mathrm{H^1}} = 1.59\cdot10^{-2}$.
These results confirm that the proposed formulation is suitable for solving design optimization problems on today's \gls{qa} hardware. We would like to stress that this direct approach does not require iterations or classical methods for the structural problem. Nevertheless, it can also be seen that the complexity of the problem is limited for practical reasons described above. In the following section, we will turn our attention to contextualizing these results within the broader landscape of \gls{qa} for design optimization.

\section{Discussion}
\label{sec:discussion}
In this study, we introduced a novel formulation for structural design optimization problems tailored to leverage \gls{qa} capabilities. Our findings show that the proposed formulation can indeed be used to solve such optimization problems and, for small-scale problems, already runs on today's \gls{qa} hardware. This demonstrates its potential for advancing computational methods in structural design optimization.
\bigskip\par
In addition to the validation of the proposed formulation, we found that it is important to account for the limitations of the \gls{qa} hardware and, where possible, develop corresponding strategies. This applies to both the number of available physical qubits and couplers and external disturbances, which can introduce noise and interfere with the quantum states. Regarding the former, we showed that specific choices in the formulation --- in this case switching from a description in stresses to an equivalent one in forces --- can significantly decrease the number of required input qubits. In addition, we have seen that the additional logical qubits and interactions due to the degree reduction technique, which is necessary in the design optimization problem, pose a challenge to the hardware by leading to longer chains that are more susceptible to disturbances. Lastly, we observed that precision issues can arise when the range of values in the problem formulation exceeds the dynamic range of the control parameters, an aspect that is closely related to the scaling and weighting of the individual terms in the problem. In this regard, we found that it is important to carefully choose the penalty weight. Furthermore, we showed that, for the structural analysis problem, it can be useful to relax the penalty weight used for \gls{qa} and add a classical local search with increased penalty weight as a post-processing step.
\bigskip\par
Comparing our results with existing literature that investigates \gls{qa} for design optimization problems, our results align with the findings that the proposed approaches, which use \gls{qa} to different extents, can be useful. However, we also find that currently only problems of limited scale and complexity are suitable to be solved by these methods due to limitations in the available hardware. In distinction to previous studies however, we were able to obtain results through an approach that solely relies on \gls{qa}, specifically without the use of classical analysis models and methods, such as the \gls{fem}, or iterative procedures.
\bigskip\par 
Although we were able to demonstrate the functionality of the formulation and find an optimal structural design, our research is not without limitations. Most notably, the application of the proposed methodology in its current form to large-scale problems is constrained by the available hardware. While the formulation proves effective for small-scale optimization problems, scaling it up struggles with a lack of robustness in its practical application. As mentioned above, this is related to the emergence of long chains during the minor-embedding, especially when the design optimization problem relies on a degree reduction technique. In addition, the limited precision in mapping the coefficients from the problem formulation onto the hardware prevents to enforce the constraint of static admissibility via a large penalty weight. 
Ways to overcome these limitations will require both advances in the capabilities of the hardware and further developments in the course of the formulation of the optimization problem.
\bigskip\par
In this sense, our findings highlight the need for further research to tackle the scalability challenges imposed by current hardware limitations. Thus, future work, on the one hand, may focus on refining the proposed formulation, e.g., by replacing the penalty term by an ansatz that fulfills the constraint of static admissibility by construction, which would address the precision issues mentioned above. On the other hand, it may also be worthwhile to explore alternative architectures for \gls{qa} that allow higher-order interactions~\cite{Lechner2015} and would obviate the need for degree reduction techniques. Finally, the development of efficient hybrid approaches represents an alternative direction that may allow to fully exploit existing hardware and extend the scope of the presented formulation towards large-scale problems.
\bigskip\par
In conclusion, our work contributes a novel formulation for structural design optimization problems to exploit the potential of \gls{qa}. While the current study can be seen as a proof of concept for our approach, the scalability limitations imposed by existing hardware necessitates further research that focuses both on theoretical advancements in the problem formulation as well as on practically relevant strategies for solving these problems on available \gls{qa} systems. Parallel to the progress made in ongoing hardware development, the evolving field of \gls{qa}-based approaches for engineering design optimization problems presents exciting opportunities for future advancements. Here, our work can lay the groundwork for continued exploration in this promising intersection of engineering optimization problems and innovative quantum-based computing approaches like \gls{qa}.


\printbibliography

@Inbook{Bendsoe2004,
    author="Bends{\o}e, Martin P.
    and Sigmund, Ole",
    title="Topology optimization by distribution of isotropic material",
    bookTitle="Topology Optimization: Theory, Methods, and Applications",
    year="2004",
    publisher="Springer Berlin Heidelberg",
    address="Berlin, Heidelberg, Germany",
    pages="1--69",
    doi="10.1007/978-3-662-05086-6_1",
}

@misc{Boothby2020,
  title={Next-generation topology of D-Wave quantum processors},
  author={Boothby, Kelly and Bunyk, Paul and Raymond, Jack and Roy, Aidan},
  doi={10.48550/arXiv.2003.00133},
  year={2020}
}

@misc{Dattani2019,
   author = {Nike Dattani},
   doi = {10.48550/arXiv.1901.04405},
   title = {Quadratization in discrete optimization and quantum mechanics},
   year = {2019},
}

@misc{DWaveLeap2023,
    title        = {D-Wave Leap Cloud Service},
    author       = {},
    year         = {},
    note         = {Accessed: 29.11.2023},
    howpublished = {\url{https://cloud.dwavesys.com/leap/}}
}

@misc{DWaveOcean2023,
    title        = {D-Wave Ocean Software Documentation},
    author       = {},
    year         = {},
    note         = {Accessed: 29.11.2023},
    howpublished = {\url{https://docs.ocean.dwavesys.com/en/stable/}}
}

@article{Engesser1889,
   author = {Friedrich Engesser},
   journal = {Zeitschrift des Architekten-und Ingenieur-Vereins zu Hannover},
   pages = {733-744},
   title = {Über statisch unbestimmte Träger bei beliebigem Formänderungs-Gesetze und über den Satz von der kleinsten Ergänzungsarbeit},
   volume = {35},
   year = {1889},
}

@misc{engioptiqa2023,
  author = {Fabian Key and Lukas Freinberegr},
  title = {EngiOptiQA (v0.1.0)},
  doi = {10.5281/zenodo.10222618}
}

@misc{FixstarsAmplify2023,
    title        = {Fixstars Amplify SDK},
    author       = {},
    year         = {},
    note         = {Accessed: 29.11.2023},
    howpublished = {\url{https://amplify.fixstars.com/en/sdk}}
}

@Inbook{Freedman2005,
   author = {D Freedman and P Drineas},
   booktitle={2005 IEEE Computer Society Conference on Computer Vision and Pattern Recognition (CVPR'05)},
   editor = {Cordelia Schmid and Stefano Soatto and Carlo Tomasi},
   doi = {10.1109/CVPR.2005.143},
   pages = {939-946},
   title = {Energy minimization via graph cuts: settling what is possible},
   volume = {2},
   year = {2005},
}

@book{Hoff1956,
   author = {Nicholas John Hoff},
   publisher = {John Wiley \& Sons},
   address = {New York, NY, USA},
   title = {The analysis of structures: based on the minimal principles and the principle of virtual displacements},
   year = {1956},
   pages = {338}
}

@article{Ishikawa2011,
   author = {H Ishikawa},
   doi = {10.1109/TPAMI.2010.91},
   journal = {IEEE Trans Pattern Anal Mach Intell},
   pages = {1234-1249},
   title = {Transformation of General Binary MRF Minimization to the First-Order Case},
   volume = {33},
   year = {2011},
}

@article{Kolmogorov2004,
   author = {V Kolmogorov and R Zabin},
   doi = {10.1109/TPAMI.2004.1262177},
   journal = {IEEE Trans Pattern Anal Mach Intell},
   pages = {147-159},
   title = {What energy functions can be minimized via graph cuts?},
   volume = {26},
   year = {2004},
}

@article{Lechner2015,
   author = {Wolfgang Lechner and Philipp Hauke and Peter Zoller},
   doi = {10.1126/sciadv.1500838},
   journal = {Sci Adv},
   month = {7},
   note = {doi: 10.1126/sciadv.1500838},
   pages = {e1500838},
   publisher = {American Association for the Advancement of Science},
   title = {A quantum annealing architecture with all-to-all connectivity from local interactions},
   volume = {1},
   year = {2015},
}

@Inbook{Mang2018,
    author="Mang, Herbert A.
    and Hofstetter, G{\"u}nter",
    title="Energieprinzipien",
    bookTitle="Festigkeitslehre",
    year="2018",
    publisher="Springer Berlin Heidelberg",
    address="Berlin, Heidelberg, Germany",
    pages="135--152",
    doi="10.1007/978-3-662-57564-2_5",
}

@article{Maruo2022,
   author = {A Maruo and T Soeda and H Igarashi},
   doi = {10.1109/TMAG.2022.3184325},
   journal = {IEEE Trans Magn},
   pages = {1-4},
   title = {Topology Optimization of Electromagnetic Devices Using Digital Annealer},
   volume = {58},
   year = {2022},
}

@article{Maruo2020,
   author = {A Maruo and H Igarashi and H Oshima and S Shimokawa},
   doi = {10.1109/TMAG.2019.2957805},
   journal = {IEEE Trans Magn},
   pages = {1-4},
   title = {Optimization of Planar Magnet Array Using Digital Annealer},
   volume = {56},
   year = {2020},
}

@article{Matsumori2022,
   author = {Tadayoshi Matsumori and Masato Taki and Tadashi Kadowaki},  
   doi = {10.1038/s41598-022-16149-8},
   journal = {Sci Rep},
   pages = {12143},
   title = {Application of QUBO solver using black-box optimization to structural design for resonance avoidance},
   volume = {12},
   year = {2022},
}

@misc{Neukart2019,
   author = {Dyon van Vreumingen and Florian Neukart and David Von Dollen and Carsten Othmer and Michael Hartmann and Arne-Christian Voigt and Thomas Bäck},
   doi = {10.48550/arXiv.1908.03947},
   title = {Quantum-assisted finite-element design optimization},
   year = {2019},
}

@article{Okada2023,
   author = {A Okada and H Yoshida and K Kidono and T Matsumori and T Takeno and T Kadowaki},
   doi = {10.1109/ACCESS.2023.3271969},
   journal = {IEEE Access},
   pages = {44343-44349},
   title = {Design Optimization of Noise Filter Using Quantum Annealer},
   volume = {11},
   year = {2023},
}

@book{Reddy2017,
   author = {Junuthula Narasimha Reddy},
   publisher = {John Wiley \& Sons},
   address = {Hoboken, USA, Chichester, UK},
   title = {Energy principles and variational methods in applied mechanics},
   edition = {Third edition},
   year = {2017},
   pages = {204-206}
}

@article{Srivastava2019,
   author = {Siddhartha Srivastava and Veera Sundararaghavan},
   doi = {10.1103/PhysRevA.99.052355},
   journal = {Phys Rev A},
   pages = {52355},
   publisher = {American Physical Society},
   title = {Box algorithm for the solution of differential equations on a quantum annealer},
   volume = {99},
   year = {2019},
}

@article{TostiBalducci2022,
   author = {Giorgio Tosti Balducci and Boyang Chen and Matthias Möller and Marc Gerritsma and Roeland De Breuker},
   doi = {10.3389/fmech.2022.914241},
   journal = {Front Mech Eng},
   title = {Review and perspectives in quantum computing for partial differential equations in structural mechanics},
   volume = {8},
   year = {2022},
}

@article{Westergaard1942,
   author = {H M Westergaard},
   doi = {10.1061/TACEAT.0005550},
   journal = {Transactions of the American Society of Civil Engineers},
   month = {1},
   pages = {765-793},
   publisher = {American Society of Civil Engineers},
   title = {On the Method of Complementary Energy},
   volume = {107},
   year = {1942},
}

@article{Wils2023,
   author = {Kevin Wils and Boyang Chen},
   doi = {10.3390/math11163451},
   journal = {Mathematics},
   title = {A Symbolic Approach to Discrete Structural Optimization Using Quantum Annealing},
   volume = {11},
   year = {2023},
}

@article{Yarkoni2022,
   author = {Sheir Yarkoni and Elena Raponi and Thomas Bäck and Sebastian Schmitt},
   doi = {10.1088/1361-6633/ac8c54},
   journal = {Rep Prog Phys},
   pages = {104001},
   publisher = {IOP Publishing},
   title = {Quantum annealing for industry applications: introduction and review},
   volume = {85},
   year = {2022},
}

@article{Ye2023,
   author = {Z Ye and X Qian and W Pan},
   doi = {10.1109/TQE.2023.3266410},
   journal = {IEEE Transactions on Quantum Engineering},
   pages = {1-15},
   title = {Quantum Topology Optimization via Quantum Annealing},
   volume = {4},
   year = {2023},
}

\end{document}